\documentclass[10pt,journal,compsoc]{IEEEtran}
\usepackage[table]{xcolor}
\usepackage{etex}
\usepackage{booktabs}
\usepackage[utf8]{inputenc}
\usepackage[ruled,vlined]{algorithm2e}
\usepackage[T1]{fontenc}
\usepackage{microtype}
\usepackage{graphicx}
\usepackage{paralist}
\usepackage{tabularx}
\usepackage{balance}
\usepackage{adjustbox}
\usepackage{multicol}
\usepackage{multirow}
\usepackage{caption}
\usepackage{pbox}
\usepackage{enumitem}
\usepackage{amssymb}
\usepackage{pifont}
\usepackage{xspace}
\usepackage{url}
\usepackage{tikz}
\usepackage{rotating}
\usepackage{float}
\usepackage[TABBOTCAP]{subfigure}
\usepackage{ragged2e}
\usepackage{listings}
\usepackage{tcolorbox}
\usepackage{booktabs}
\usepackage{fixltx2e}
\usepackage[framemethod=TikZ]{mdframed}
\usepackage{lipsum}

\definecolor{Gray}{gray}{0.9}
\definecolor{codegreen}{rgb}{0,0.6,0}
\definecolor{codegray}{rgb}{0.73,0.38,0.06}
\definecolor{codepurple}{rgb}{0.70,0.27,0}
\definecolor{codemagenta}{rgb}{0.74,0.09,0.42}
\definecolor{codeoutput}{rgb}{0.5,0,0}
\definecolor{backcolour}{rgb}{0.96,0.96,0.96}

\ifCLASSOPTIONcompsoc
  \usepackage[nocompress]{cite}
\else
  \usepackage{cite}
\fi

  {\list{}{\leftmargin=0.1in\rightmargin=0.1in}\item[]}%
  {\endlist}

\newcommand{\ie}{\emph{i.e.,}\xspace}
\newcommand{\eg}{\emph{e.g.,}\xspace}
\newcommand{\etc}{etc.\xspace}
\newcommand{\etal}{\emph{et~al.}\xspace}
\newcommand{\secref}[1]{Section~\ref{#1}\xspace}

\newcommand{\figref}[1]{Fig.~\ref{#1}\xspace}

\newcommand{\tabref}[1]{Table~\ref{#1}\xspace}

\newcommand{\ReportDataset}[2]{$\mathit{#1}_{\mathit{#2}}$\xspace}

\newcommand{\BF}[1]{\ReportDataset{BF}{#1}}
\newcommand{\BFsmall}{\BF{small}}
\newcommand{\BFmedium}{\BF{medium}}

\newcommand{\AG}[1]{\ReportDataset{AG}{#1}}
\newcommand{\AGabs}{\AG{abs}}
\newcommand{\AGraw}{\AG{raw}}

\newcommand{\MG}[1]{\ReportDataset{MG}{#1}}
\newcommand{\MGident}{\MG{ident}}

\newcommand{\CS}{\ReportDataset{CS}{}}
\newcommand{\rev}[1]{\textcolor{black}{#1}}

\newboolean{showcomments}

\setboolean{showcomments}{true}

\ifthenelse{\boolean{showcomments}}
  {\newcommand{\nb}[2]{
    \fbox{\bfseries\sffamily\scriptsize#1}
    {\sf\small$\blacktriangleright$\textit{#2}$\blacktriangleleft$}
   }
  }
  {\newcommand{\nb}[2]{}
  }

\usepackage{inconsolata}
\lstdefinestyle{mystyle}{
	backgroundcolor=\color{backcolour},   
	commentstyle=\color{codegreen},
	keywordstyle=\color{codepurple},
	numberstyle=\tiny\color{codegray},
	stringstyle=\color{codemagenta},
	language=Java,
	breakatwhitespace=false,         
	breaklines=true,                 
	keepspaces=true,                 
	numbers=none,                    
	numbersep=5pt,                  
	showspaces=false,                
	showstringspaces=false,
	showtabs=false,                  
	tabsize=2,
	frame=tb,
	framerule=0pt,
	basicstyle=\fontsize{5.5}{5.5}\fontfamily{\ttdefault}\selectfont
}
\lstdefinestyle{mystyleresult}{
	backgroundcolor=\color{backcolour},   
	commentstyle=\color{codegreen},
	keywordstyle=\color{codeoutput},
	numberstyle=\tiny\color{codegray},
	stringstyle=\color{red},
	language=Java,
	breakatwhitespace=false,         
	breaklines=true,                 
	keepspaces=true,                 
	numbers=none,                    
	numbersep=5pt,                  
	showspaces=false,                
	showstringspaces=false,
	showtabs=false,                  
	tabsize=2,
	frame=tb,
	framerule=0pt,
	basicstyle=\color{codeoutput}\fontsize{5.5}{5.5}\fontfamily{\ttdefault}\selectfont
}

\begin{document}

\title{Using Transfer Learning for Code-Related Tasks}

\author{Antonio~Mastropaolo,
	Nathan~Cooper,
	David~Nader~Palacio,
	Simone~Scalabrino,\\
	Denys~Poshyvanyk,
	Rocco~Oliveto,
	and~Gabriele~Bavota
	\IEEEcompsocitemizethanks{\IEEEcompsocthanksitem A. Mastropaolo is with SEART @  Software Institute, Universit\`a della Svizzera italiana, Switzerland. \protect\\
		E-mail: antonio.mastropaolo@usi.ch
		\IEEEcompsocthanksitem N. Cooper is with SEMERU @ William \& Mary, USA. \protect\\
		E-mail: nacooper01@email.wm.edu
		\IEEEcompsocthanksitem D. Nader~Palacio is with SEMERU @ William \& Mary, USA. \protect\\
		E-mail: danaderp@gmail.com
		\IEEEcompsocthanksitem S. Scalabrino is with University of Molise, Italy. \protect\\
		E-mail: simone.scalabrino@unimol.it
		\IEEEcompsocthanksitem D. Poshyvanyk is with SEMERU @ William \& Mary, USA. \protect\\
		E-mail: denys@cs.wm.edu
		\IEEEcompsocthanksitem R. Oliveto is with University of Molise, Italy. \protect\\
		E-mail: rocco.oliveto@unimol.it
		\IEEEcompsocthanksitem G. Bavota is with SEART @  Software Institute, Universit\`a della Svizzera italiana, Switzerland. \protect\\
	E-mail: gabriele.bavota@usi.ch}
}

\markboth{Journal of \LaTeX\ Class Files,~Vol.~xx, No.~x, Month~xxxx}%
{Mastropaolo \MakeLowercase{\etal}: Using Transfer Learning for Code-Related Tasks}
\IEEEtitleabstractindextext{%
\begin{abstract}
	Deep learning (DL) techniques have been used to support several code-related tasks such as code summarization and bug-fixing. In particular, pre-trained transformer models are on the rise, also thanks to the excellent results they achieved in Natural Language Processing (NLP) tasks. The basic idea behind these models is to first pre-train them on a generic dataset using a self-supervised task (\eg filling masked words in sentences). Then, these models are fine-tuned to support specific tasks of interest (\eg language translation). A single model can be fine-tuned to support multiple tasks, possibly exploiting the benefits of \emph{transfer learning}. This means that knowledge acquired to solve a specific task (\eg language translation) can be useful to boost performance on another task (\eg sentiment classification). While the benefits of transfer learning have been widely studied in NLP, limited empirical evidence is available when it comes to code-related tasks. In this paper, we assess the performance of the Text-To-Text Transfer Transformer (T5) model in supporting four different code-related tasks: (i) automatic bug-fixing, (ii) injection of code mutants, (iii) generation of assert statements, and (iv) code summarization. We pay particular attention in studying the role played by pre-training and multi-task fine-tuning on the model's performance. We show that (i) the T5 can achieve better performance as compared to state-of-the-art baselines; and (ii) while pre-training helps the model, not all tasks benefit from a multi-task fine-tuning. 
\end{abstract}

	}

\maketitle

\IEEEdisplaynontitleabstractindextext
\IEEEpeerreviewmaketitle

\section{Introduction} \label{sec:intro}

Several code-related tasks have been recently automated by researchers exploiting Deep Learning (DL) techniques \cite{watson2020systematic}. Several of these works customize DL models proposed in the Natural Language Processing (NLP) field to support code-related tasks, and most of them share one common characteristic: \emph{They shape the problem at hand as a text-to-text transformation, in which the input and the output of the model are text strings}. For instance, Tufano \etal \cite{tufan2021towards} used an encoder-decoder architecture, commonly adopted in Neural Machine Translation (NMT) \cite{Kalchbrenner:2013,Sutskever:2014,Cho:2014}, to predict code changes usually recommended by reviewers in a code review process. Both the input and output are represented as a stream of tokens (\ie textual format), with the input being the code submitted for review and the output a revised code implementing changes likely to be required in the code review process. While this is only one concrete example, similar observations hold for techniques automating bug fixing \cite{Tufano:tosem2019, Chen:2019,Mesbah:fse2019,Hata:2018},
learning generic code changes \cite{Tufano:icse2019}, supporting code migration \cite{Nguyen:icse2014,Nguyen:fse2013}, code summarization \cite{LeClair:icse2019,Jiang:ASE'17,Liu:ase2018,haque:2020}, code reviews \cite{tufan2021towards,tufan2022towards2}, pseudo-code generation \cite{Oda:ase2015}, code deobfuscation \cite{Vasilescu:fse2017,Jaffe:icpc2018}, injection of code mutants \cite{Tufano:icsme2019}, generation of assert statements \cite{Watson:icse2020}, clone detection \cite{White2016clones,ClonesMSR18}, traceability \cite{MoranICSETraceability} and code completion \cite{Karampatsis:DLareBest,alon2019structural,kim2020code,svyatkovskiy2020intellicode,brody2020neural, ciniselli2021empirical, ciniselli2021empirical,White:MSR15}.

Recent years have seen the rise of \emph{transfer learning} in the field of natural language processing. The basic idea is to first pre-train a model on a large and generic dataset by using a self-supervised task, \eg masking tokens in strings and asking the model to guess the masked tokens. Then, the trained model is fine-tuned on smaller and specialized datasets, each one aimed at supporting a specific task. In this context, Raffel \etal \cite{raffel2019exploring} proposed the T5 (Text-To-Text Transfer Transformer) model, pre-trained on a large natural language corpus and fine-tuned to achieve state-of-the-art performance on many tasks, all characterized by text-to-text transformations.

In our recent work \cite{mastropaolo2021empirical} we empirically investigated the potential of a T5 model when pre-trained and fine-tuned to support four code-related tasks also characterized by text-to-text transformations. In particular, we started by pre-training a T5 model using a large dataset consisting of 499,618 English sentences and 1,569,889 source code components (\ie Java methods). Then, we fine-tuned the model using four datasets from previous work with the goal of supporting four code-related tasks:

\emph{Automatic bug-fixing.} We used the dataset by Tufano \etal \cite{Tufano:tosem2019}, composed of instances in which the ``input string'' is represented by a buggy Java method and the ``output string'' is the fixed version of the same method.

\emph{Injection of code mutants.} This dataset is also by Tufano \etal \cite{Tufano:icsme2019}, and features instances in which the input-output strings are reversed as compared to automatic bug-fixing (\ie the input is a fixed method, while the output is its buggy version). The model must learn how to inject bugs (mutants) in code instead of fixing bugs.

\emph{Generation of assert statements in test methods.} We used the dataset by Watson \etal \cite{Watson:icse2020}, composed of instances in which the input string is a representation of a test method without an assert statement and a focal method it tests (\ie the main production method tested), while the output string encodes an appropriate assert statement for the input test method.

\emph{Code Summarization.} We used the dataset by Haque \etal \cite{haque:2020} where input strings are some representations of a Java method to summarize, \& an output string is a textual summary.

We fine-tuned a single pre-trained T5 model in a multi-task setting on all four tasks, and showed that it is able to achieve better results as compared to the four referenced baselines in all tasks \cite{Tufano:tosem2019,Watson:icse2020,haque:2020,Tufano:icsme2019}. However, since we only experimented with a pre-trained model fine-tuned in a multi-task setting, questions about the actual advantage offered by transfer learning remained unanswered. \rev{In this work, we aim at overcoming such a limitation that is also typical of several other works in the literature using off-the-shelf pre-trained models like T5 to support code related tasks (\eg \cite{Rahmani:lang,Sarim:2019}). Indeed, little effort has been spent on understanding the actual benefits (if any) that transfer learning brings when dealing with code-related tasks. Such observation holds for both (i) the pre-training phase, that should provide the model with general knowledge about a language of interest (\eg Java) being at the core of the tasks to automate (\eg bug-fixing); and (ii) the multi-task fine-tuning, that should allow the model to exploit knowledge acquired when trained for a specific task (\eg bug-fixing) also for the automation of other tasks (\eg generation of assert statements), thus possibly boosting the overall performance in all the tasks. Besides the expected positive impact on performance, pre-training and multi-task fine-tuning are also useful in real-life scenarios in which the training data for a particular task of interest is scarce (\eg when manually labeled instances are needed) \cite{Robbes:icse2019}. Pre-training the model in an unsupervised setting and/or fine-tuning it on other related tasks for which more training data is available can unlock the possibility of using deep learning models also for tasks characterized by scarcity of training data.}

In this paper, we extend our previous work \cite{mastropaolo2021studying} by carefully assessing the impact of both pre-training and multi-task fine-tuning on the T5 performance. In particular, we assess the performance of the T5 in the following scenarios:

\begin{itemize}
	\item \textbf{No Pre-training:} We do not run any pre-training step. We directly fine-tune four different T5 models, each one supporting one of the four tasks we experiment with.
	\smallskip
	\item\textbf{Pre-training single task:} We first pre-train the T5 model on the dataset presented in \tabref{tab:pretrain_dataset}. Then, starting from it, we fine-tune four models, one for each single task.
	\smallskip
	\item\textbf{Pre-training Multi-Task}: Lastly, we fine-tune the pre-trained model using a multi-task learning framework in which we train a single model to support all four code-related tasks. We experiment with two different multi-task fine-tunings: (i) the first is the one used in our original paper \cite{mastropaolo2021studying}, in which the percentage of training instances from each of the four tasks is proportional to the size of their training dataset; (ii) the second in which the percentage of training instances is the same for all four tasks (\ie 25\% per task). 
\end{itemize}

In total, this resulted in the training, hyperparameters tuning, and testing of ten different models.  \rev{Note that the choice of the four tasks subject of our study (\ie \emph{bug-fixing}, \emph{mutants injection}, \emph{asserts generation}, and \emph{code summarization}) is dictated by the will of experimenting with tasks that use, represent, and manipulate code in different ways. In particular, we include in our study tasks aimed at (i) transforming the input code with the goal of changing its behavior (\emph{bug-fixing} and \emph{mutants injection}); (ii) ``comprehending code'' to verify its behavior (\emph{asserts generation}); and (iii) ``comprehending code'' to summarize it in natural language (\emph{code summarization}). Also, following what has been done in the original datasets from previous work, the four tasks involve abstracted source code (\emph{bug-fixing} \cite{Tufano:tosem2019}, \emph{mutants injection} \cite{Tufano:icsme2019}, and \emph{asserts generation} \cite{Watson:icse2020}), raw source code (\emph{asserts generation} \cite{Watson:icse2020} and \emph{code summarization} \cite{haque:2020}), and natural language (code summarization \cite{haque:2020}). Such a mix of tasks helps in increasing the generalizability of our findings.}

We also perform a novel analysis of our dataset aimed at assessing the generalizability of our models by looking at the level of data snooping among our training and test datasets.

Our results confirm that the T5 can substantially boost the performance on all four code-related tasks. For example, when the T5 model is asked to generate assert statements on raw source code, $\sim$70\% of test instances are successfully predicted by the model, against the 18\% of the original baseline \cite{Watson:icse2020}. Also, we show that the pre-training is beneficial for all tasks, while the multi-task fine-tuning does not consistently help in improving performance. Finally, our datasets analysis confirm the generalizability of the tested models.
The code and data used in this work are publicly available \cite{replication}.

 
\section{Background and Related Work} \label{sec:related}

In recent years, DL techniques have been increasingly used to support software engineering (SE). The activities commonly supported by state-of-the-art approach include software maintenance and software testing \cite{yang2020survey}, and most of the proposed approaches target the source code \cite{watson2020systematic}. While available approaches support a plethora of concrete SE tasks \cite{yang2020survey, watson2020systematic}, in this section we focus on the ones we target in our study: automated bug-fixing, injection of code mutants, generation of assert statements in test methods, and code summarization. We discuss in detail the techniques we use as baselines for each task. A broader literature review on the topic is available in two recent surveys by Yang \etal \cite{yang2020survey} and Watson \etal \cite{watson2020systematic}.

\subsection{Automatic Bug-Fixing}
Many techniques have been proposed for the automatic fixing of software bugs. Several of them \cite{LeGoues:tse2012,LeGoues:icse2012,Sidiroglou-Douskos:2015,Pierret:2009,Gabel:2010,Carzaniga:2013,Nguyen:2013:ASE,White:2019:SANER,Bader:oopsla2019} rely on the \emph{redundancy assumption}, claiming that large programs contain the seeds of their own repair. Such an assumption has been verified by at least two independent studies~\cite{Martinez:2014,Barr:2014}. 
Automated bug-fixing techniques based on DL can rely on different levels of code abstraction. Word tokenization is a commonly used one, even if higher-level abstractions (\eg AST-based) allow to achieve better results \cite{namavar2021controlled}.

Mesbah \etal \cite{Mesbah:fse2019} focus on build-time compilation failures by presenting DeepDelta, an approach using NMT to fix the build. The input is represented by features characterizing the compilation failure (\eg kind of error, AST path, \etc). As output, DeepDelta provides the AST changes needed to fix the error. In the presented empirical evaluation, DeepDelta correctly fixed 19,314 out of 38,788 (50\%) compilation errors.

Chen \etal \cite{Chen:2019} present SequenceR, a sequence-to-sequence approach trained on over 35k single-line bug-fixes. SequenceR takes as input the buggy line together with relevant code lines from the buggy class (\textit{abstract buggy context}). The output of the approach is the recommended fix for the buggy line. The approach, tested on a set of 4,711 bugs, was able to automatically fix 950 ($\sim$20\%) of them. Similar approaches have been proposed by Hata \etal \cite{Hata:2018} and Tufano \etal \cite{Tufano:tosem2019}. The latter is the one we compared our approach with and, thus, we describe it in more details.

\begin{figure}[h]
	\centering
	\includegraphics[width=0.65\linewidth]{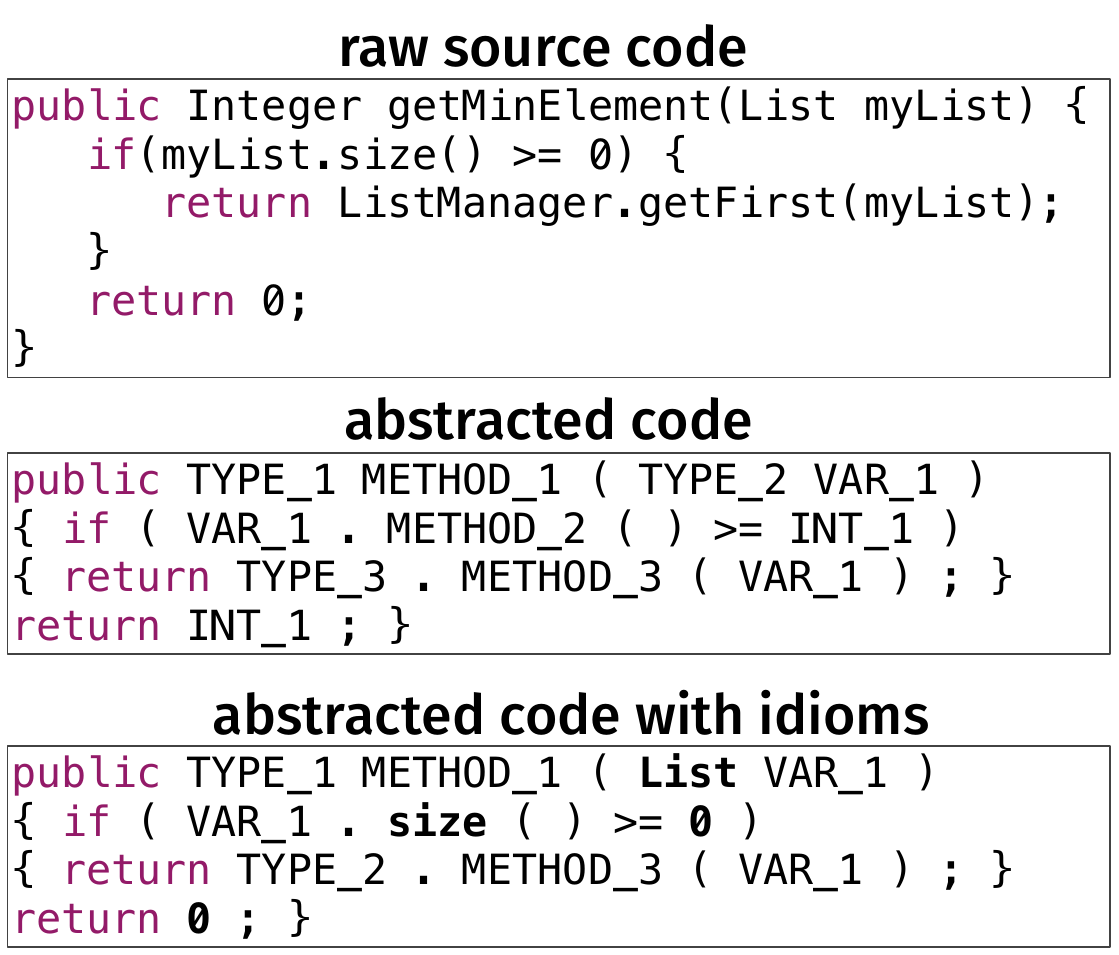}
	\caption{Abstraction process \cite{Tufano:tosem2019}}
	\label{fig:abstraction}
\end{figure}

Tufano \etal \cite{Tufano:tosem2019} investigate the performance of an NMT-based approach in the context of automatic bug-fixing. 
They train an encoder-decoder model on a set of bug-fix pairs (BFPs), meaning pairs of strings in which the first one (input) represents a Java method that has been subject to a bug-fixing activity, and the second one (target) represents the same Java method once the bug was fixed. 
To build this dataset, the authors mined $\sim$787k bug-fixing commits from GitHub, from which they extracted $\sim$2.3M BFPs. After that, the code of the BFPs is abstracted to make it more suitable for the NMT model (\ie to reduce the vocabulary of terms used in the source code identifiers and literals).  The abstraction process is depicted in \figref{fig:abstraction}. 

The top part of the figure represents the raw source code to abstract. The authors use a Java lexer and a parser to represent each method as a stream of tokens, in which Java keywords and punctuation symbols are preserved and the role of each identifier (\eg whether it represents a variable, method, \etc) as well as the type of a literal is discerned. 

IDs are assigned to identifiers and literals by considering their position in the method to abstract: The first variable name found will be assigned the ID of VAR\_1, likewise the second variable name will receive the ID of VAR\_2. This process continues for all identifiers as well as for the literals (\eg STRING\_X, INT\_X, FLOAT\_X). The output of this stage is the code reported in the middle of \figref{fig:abstraction} (\ie abstracted code). Since some identifiers and literals appear very often in the code (\eg variables \texttt{i}, \texttt{j}, literals \texttt{0}, \texttt{1}, method names such as \texttt{size}), those are treated as ``idioms'' and are not abstracted (see bottom part of \figref{fig:abstraction}, idioms are in bold). Tufano \etal consider as idioms the top $0.005\%$ frequent words in their dataset. During the abstraction a mapping between the raw and the abstracted tokens is maintained, thus allowing to reconstruct the concrete code from the abstract code generated by the model. 

The set of abstracted BFPs has been used to train and test the approach. The authors build two different sets, namely $BFP_{small}$, only including methods having a maximum length of 50 tokens (for a total of 58,350 instances), and $BFP_{medium}$, including methods up to 100 tokens (65,455). The model was able to correctly predict the patch for the buggy code in 9\% and 3\% of cases in the $BFP_{small}$ and $BFP_{medium}$ dataset, respectively. 

While other works have tackled the automatic bug-fixing problem, the approach by Tufano \etal has been tested on a variety of different bugs, rather than on specific types of bugs/warnings (\eg only single-line bugs are considered in \cite{Chen:2019}, while compilation failures are addressed in \cite{Mesbah:fse2019}). 

Thus, we picked it as representative DL technique for automatic bug-fixing and we use the two datasets by Tufano \etal \cite{Tufano:tosem2019} to fine-tune the T5 model for the ``automatic bug-fixing'' problem, comparing the achieved performance with the one reported in the original paper. 

\subsection{Injection of Code Mutants}

Brown \etal \cite{Brown:fse2017} were the first to propose a data-driven approach for generating code mutants, leveraging bug-fixes performed in software systems to extract syntactic-mutation patterns from the diffs of patches. Tufano \etal \cite{Tufano:icsme2019} built on this concept by presenting an approach using NMT to inject mutants that mirror real bugs. The idea is to reverse the learning process used for fixing bugs \cite{Tufano:tosem2019}: The model is trained to transform correct methods (\ie the method obtained after the bug-fixing activity) into buggy methods (before the bug-fix). Indeed, the methodology used by the authors is the same used for the bug-fixing task (previously described), including the abstraction process.

This is, to date, the only DL-based technique for injecting code mutants. Thus, we use the dataset exploited by Tufano \etal \cite{Tufano:icsme2019} to fine-tune the T5 model for the problem of ``injecting code mutants'', comparing the achieved results with the ones reported in the original paper. Specifically, we reused their largest dataset, referred to as $GM_{ident}$ in the paper\footnote{A subset of this dataset named $GM_{ident-lit}$ has also been used in the original paper \cite{Tufano:icsme2019} to avoid including in the study bugs requiring the generation of previously unseen literals. We decided to test the T5 model on the most complex and complete dataset.}, featuring 92,476 training instances, 11,560 used for hyperparameter tuning (evaluation set), and 11,559 used for testing. On this data, the approach by Tufano \etal was able to correctly predict the bug to inject in 17\% of cases (1,991).

\subsection{Generation of Assert Statements in Test Methods}
Watson \etal \cite{Watson:icse2020} start from the work by Shamshiri \etal~\cite{Shamshiri:FSE'15}, who observed that tools for the automatic generation of test cases such as Evosuite \cite{evosuite}, Randoop \cite{randoop} and Agitar \cite{agitar} exhibit insufficiencies in the automatically generated assert statements. 

Thus, they propose ATLAS, an approach for generating syntactically and semantically correct unit test assert statements using NMT. To train ATLAS, the authors mined 2.5M test methods from GitHub with their corresponding {\tt assert} statement. For each of those test methods, they also identified the focal method, meaning the main production code method exercised by the test. A preprocessing of the dataset has been performed to remove all test methods longer than 1K tokens. Also, test methods requiring the synthesis of one or more unknown tokens for generating the appropriate assert statements have been removed. Indeed, if the required tokens cannot be found in the vocabulary of the test method they cannot be synthesized when the model attempts to generate the prediction. Finally, all duplicates have been removed from the dataset, leading to a final set of 158,096 Test-Assert Pairs (TAPs). Each method left in the dataset has then been abstracted using the same approach previously described by Tufano \etal \cite{Tufano:tosem2019}. However, in this case the authors experiment with two datasets, one containing raw source code and one abstracted code. ATLAS was able to generate asserts identical to the ones written by developers in 31.42\% of cases (4,968 perfectly predicted assert statements) when only considering the top-1 prediction, and 49.69\% (7,857) when looking at the top-5 in the abstracted dataset, while performance is lower on the raw dataset (17.66\% for top-1 and 23.33\% for top-5).

We use the datasets by Watson \etal \cite{Watson:icse2020} to fine-tune our T5 model for the ``generation of assert statements'' problem, and compare the achieved performance with the one in the original paper. 
Recently, Tufano \etal \cite{tufano2020generating} proposed an approach based on transformers to achieve a the same goal. Their results show that such an approach achieves better results than ATLAS \cite{Watson:icse2020}. We did not use the approach proposed by Tufano \etal \cite{tufano2020generating} as the main baseline because it is very similar to the one we presented in the our conference paper that this paper extends \cite{mastropaolo2021studying}. 

\subsection{Code Summarization}
Code summarization is one of the mainstream methods for automatic documentation of source code. The proposed summarization techniques fall into two categories. \textit{Extractive} summarization techniques generate summaries by extracting information from the code components being summarized \cite{Haiduc:wcre2010,Sridhara:icpc2011,Moreno:icpc2013,Rodeghero:icse17}. On the other hand, \textit{abstractive} summarization techniques aim at including in the summaries information not directly available in the source code \cite{Sridhara:icse2011,Burney:tse2016,Jiang:ASE'17,Hu:icpc2018,haque:2020}. DL techniques have been used to support for the latter.

Hu \etal \cite{Hu:icpc2018} use a Deep Neural Network (DNN) to automatically generate comments for a given Java method. The authors mine $\sim$9k Java projects hosted on GitHub to collect pairs of $\langle$method, comment$\rangle$, where ``comment'' is the first sentence of the Javadoc linked to the method. These pairs, properly processed, are used to train and test the DNN. The authors assess the effectiveness of their technique by using the BLEU-4 score \cite{Papineni:2002}, showing the superiority of their approach with respect to the competitive technique presented in \cite{iyer:acl}.

Allamanis \etal \cite{Allamanis:2016} use attention mechanisms in neural networks to suggest a descriptive method name starting from an arbitrary snippet of code. Their approach can name a code snippet exactly as a developer would do in $\sim$25\% of cases. 

LeClair \etal \cite{LeClair:icse2019} present a neural model combining the AST source code structure and words from code to generate coherent summaries of Java methods. The approach, tested on 2.1M methods, showed its superiority as compared to the previous works by Hu \etal \cite{Hu:icpc2018} and Iyer \etal \cite{iyer:acl}.

The approach by Haque \etal \cite{haque:2020} is the most recent in the area of DL-aided source code summarization, and it is an improvement of the work by LeClair \etal \cite{LeClair:icse2019}. 

It still aims at documenting Java methods through an encoder-decoder architecture but, in this case, three inputs are provided to the model to generate the summary: (i) the source code of the method, as a flattened sequence of tokens representing the method; (ii) its AST representation; and (iii) the ``file context'', meaning the code of every other method in the same file. The authors show that adding the contextual information as one of the inputs substantially improves the BLEU score obtained by deep learning techniques. The dataset used in the evaluation is composed of 2.1M Java methods paired with summaries. We reuse this dataset for the fine-tuning of the T5 model for the code summarization problem, and compare its performance to the state-of-the-art approach proposed by Haque \etal \cite{haque:2020}.

\section{Text-to-Text-Transfer-Transformer} \label{sec:t5}

The T5 model has been introduced by Raffel \etal \cite{raffel2019exploring} to support multitask learning in Natural Language Processing (NLP). The idea is to reframe NLP tasks in a unified text-to-text format in which the input and output are always text strings. For example, a single model can be trained to translate across languages and to autocomplete sentences. This is possible since both tasks can be represented in a text-to-text format (\eg in the case of translation, the input is a sentence in a given language, while the output is the translated sentence). T5 is trained in two phases: \textit{pre-training}, which allows defining a shared knowledge-base useful for a large class of sequence-to-sequence tasks (\eg guessing masked words in English sentences to learn about the language), and \textit{fine-tuning}, which specializes the model on a specific downstream task (\eg learning the translation of sentences from English to German). We briefly overview the T5 model and explain how we pre-trained and fine-tuned it to support the four said code-related tasks. Finally, we describe the  decoding strategy for generating the predictions.

\subsection{An Overview of T5}
T5 is based on the transformer model architecture that allows handling a variable-sized input using stacks of self-attention layers. When an input sequence is provided, it is mapped into a sequence of embeddings passed into the encoder. The T5, in particular, and a transformer model \cite{vaswani2017attention}, in general, offer two main advantages over other state-of-the-art models: (i) it is more efficient than RNNs since it allows to compute the output layers in parallel, and (ii) it is able to detect hidden and long-ranged dependencies among tokens, without assuming that nearest tokens are more related than distant ones. This last property is particularly relevant in code-related tasks since a variable declaration may be distant from its usage.
Five different versions of T5 have been proposed \cite{raffel2019exploring}: \textit{\textit{small}}, \textit{\textit{base}}, \textit{\textit{large}}, \textit{\textit{3 Billion}}, and \textit{\textit{11 Billion}}. These variants differ in terms of complexity, with the smaller model (T5\textsubscript{\textit{small}}) having 60M parameters against the 11B of the largest one (T5\textsubscript{\textit{11B}}). As acknowledged by the authors \cite{raffel2019exploring}, even if the accuracy of the most complex variants is higher than the less complex models, the training complexity increases with the number of parameters. Considering the available computational resources, we decided to use the simplest T5\textsubscript{\textit{small}} model.

\textbf{T5\textsubscript{\textit{small}} architectural details.}
The T5\textsubscript{\textit{small}} architecture is characterized by six blocks for encoders and decoders. The feed-forward networks in each block consist of a dense layer with an output dimensionality ($d_{ff}$) of 2,048. The \textit{key} and \textit{value} matrices of all attention mechanisms have an inner dimensionality ($d_{kv}$) of 64, and all attention mechanisms have eight heads. All the other sub-layers and embeddings have a dimensionality ($d_{model}$) of 512.

\subsection{Pre-training of T5}\label{subsec:training_strategy}

In the \textit{pre-training} phase we use a self-supervised task similar to the one used by Raffel \etal \cite{raffel2019exploring}, consisting of masking tokens in natural language sentences and asking the model to guess the masked tokens. However, we did not perform the pre-training by only using natural language sentences, since all the tasks we target involve source code. We use a dataset composed of both (technical) natural language (\ie code comments) and source code. To obtain the dataset for the pre-training we start from the CodeSearchNet dataset \cite{husain2019codesearchnet} which provides 6M functions from open-source code. We only focus on the $\sim$1.5M methods written in Java, since the four tasks we aim at supporting are all related to Java code and work at method-level granularity (\eg fixing a bug in a method, generating the summary of a method, \etc). 

Then, since for three of the four tasks we support (\ie \emph{automatic bug-fixing} \cite{Tufano:tosem2019}, \emph{generation of assert statements} \cite{Watson:icse2020}, and \emph{injection of code mutants} \cite{Tufano:icsme2019}) the authors of the original papers used an abstracted version of source code (see \secref{sec:related}), we used the {\tt src2abs} tool by Tufano \cite{Tufano:tosem2019} to create an abstracted version of each mined Java method. In the abstraction process, special tokens are used to represent identifiers and literals of the input method. For example, the first method name found (usually the one in the method signature) will be assigned the \texttt{METHOD\_1} token, likewise the second method name (\eg a method invocation) will be represented by \texttt{METHOD\_2}. This process continues for all the method and variable names (\texttt{VAR\_X}) as well as the literals (\texttt{STRING\_X}, \texttt{INT\_X}, \texttt{FLOAT\_X}). Basically, the abstract method consists of language keywords (\eg \texttt{for}, \texttt{if}), separators (\eg ``('', ``;'', ``\}'') and special tokens representing identifiers and literals. Comments and annotations are removed during abstraction. Note that, since the tool was run on Java methods in isolation (\ie without providing it the whole code of the projects they belong to), {\tt src2abs} raised a parsing error in $\sim$600k of the $\sim$1.5M methods (due \eg to missing references), leaving us with $\sim$900k abstracted methods. We still consider such a dataset as sufficient for the pre-training. 

The CodeSearchNet dataset does also provide, for a subset of the considered Java source code methods, the first sentence in their Javadoc. We extracted such a documentation using the {\tt docstring\_tokens} field in CodeSearchNet, obtaining it for 499,618 of the considered methods. We added these sentences to the pre-training dataset. This whole process resulted in a total of 2,984,627 pre-training instances, including raw source code methods, abstracted methods, and code comment sentences. In the obtained dataset there could be duplicates between (i) different raw methods that become equal once abstracted, and (ii) comments re-used across different methods. Thus, we remove duplicates, obtaining the final set of 2,672,423 instances reported in \tabref{tab:pretrain_dataset}. This is the dataset we use for pre-training the T5 model, using the BERT-style objective function Raffel \etal used in their experiments and consisting of randomly masking 15\% of tokens (\ie words in comments and code tokens in the raw and abstracted code). 

\begin{table}[h!]
		\caption{Datasets used for the pre-training of T5.}
	\centering
	\begin{tabular}{lr}
		\toprule
		\textbf{Data sources}                        & \textbf{Instances}  \\
		\midrule
		Source code                  & 1,569,773           \\
		Abstracted source code        &   766,126           \\
		Technical natural language    &   336,524           \\
		\midrule                                     
		\textbf{Total} & 2,672,423          \\
		\bottomrule
	\end{tabular}
	\label{tab:pretrain_dataset}
\end{table}

Finally, since we pre-train and fine-tune the models on a software-specific dataset, we create a new \emph{SentencePiece} model \cite{DBLP:journals/corr/abs-1808-06226} (\ie a tokenizer for neural text processing) by training it on the entire pre-training dataset so that the T5 model can properly handle the \emph{Java} language and its abstraction. This model implements subword units (\eg byte-pair-encoding BPE)  and unigram language model \cite{kudo2018subword} to alleviate the open vocabulary problem in neural machine translation. The pre-training of the models has been performed for 250k steps which, using a batch size of 128 results in $\sim$32M of masked code instances processed that, given the size of the pre-training dataset (see \tabref{tab:pretrain_dataset}) correspond to $\sim$12 epochs.


\subsection{Fine-tuning of T5}\label{subsec:finetuning}
We detail the process used to fine-tune the T5 model.
\rev{Before explaining how the training instances are represented within each fine-tuning dataset, it is important to clarify that both in the pre-training and in the fine tuning the T5 can handle any sort of training instance as long as it can be formulated as a text-to-text transformation. Indeed, the T5 represents each training dataset as a $N\times2$ matrix, where $N$ is the number of instances in the dataset and the 2 dimensions allow to express the input text and the expected output text. In the case of pre-training, the input text is an instance (\ie a raw method, an abstract method, or a Javadoc comment) in which 15\% of tokens have been masked, while the output text represents the correct predictions for the masked tokens. In the four downstream tasks, instead, the text-to-text pairs are represented as explained in the following.}

\subsubsection{Fine-tuning dataset} \label{sub:finetuning_datasets}
We describe the datasets we use for \textit{fine-tuning} the model for the four targeted tasks. The datasets are summarized in \tabref{tab:finetuning_datasets}. The number of training steps performed for the different tasks is proportional to the size of their training dataset. Indeed, we aim at ensuring that the same number of ``epochs'' is performed on each training dataset. Thus, smaller training datasets require a lower number of steps to reach the same number of epochs of larger datasets. In particular, we used 1.75M fine-tuning steps for the \emph{multi-task} setting $\sim$90 epochs) and we scaled the others proportionally to reach the same number of epochs (\eg $\sim$1.41M for the \emph{code summarization} task).

\begin{table*}[t]
	\centering
	\caption{Task-specific datasets used for fine-tuning T5.}
		\begin{tabular}{llrrr}
			\toprule
			\textbf{Task}  & \textbf{Dataset}                & \textbf{Training-set}  & \textbf{Evaluation-set}  & \textbf{Test-set} \\
			\midrule                                          
			\multirow{2}{*}{Automatic Bug-Fixing}
			& \BFsmall \cite{Tufano:tosem2019}  & 46,680 & 5,835                                   & 5,835             \\      
			& \BFmedium \cite{Tufano:tosem2019} & 52,364  & 6,546                                  & 6,545             \\
			\midrule
			Injection of Code Mutants
			& \MGident \cite{Tufano:icsme2019}  & 92,476 & 11,560                                 & 11,559            \\
			\midrule
			\multirow{2}{*}{Generation of Asserts in Test} 
			& \AGabs \cite{Watson:icse2020}     & 126,477 & 15,809                                 & 15,810            \\
			& \AGraw \cite{Watson:icse2020}     & 150,523 & 18,816                                 & 18,815            \\
			\midrule
			Code Summarization
			& \CS \cite{haque:2020}             & 1,953,940 & 104,272                              & 90,908            \\
			
			\midrule                                          
			\textbf{Total} &                             & 2,422,460     & 162,838                              & 149,472           \\
			\bottomrule
		\end{tabular}
	
	\label{tab:finetuning_datasets}
\end{table*}

\textbf{Automatic Bug Fixing (BF)}. 
We use the dataset by Tufano \etal \cite{Tufano:tosem2019} composed by triplets $\mathit{BF}_m = \langle  m_\mathit{b}, m_\mathit{f}, M \rangle$, where $m_\mathit{b}$ and $m_\mathit{f}$ are the abstracted version of the buggy and fixed version of Java method, respectively, and $M$ represents the mapping between the abstracted tokens and the raw code tokens (\eg \texttt{VAR\_1} $\rightarrow$ \texttt{webServerPort}), which allows to track back the output of the model to source code.
The triplets refer to methods with at most 100 tokens and they are split into two sub-datasets: (i) the \textit{small} version, containing methods with up to 50 tokens, and a \textit{medium} version, with methods with at most 100 tokens.
We train the model to predict the fixed versions, $m_\mathit{f}$, given the buggy versions, $m_\mathit{b}$. Given the presence of two datasets, we divide the BF task in two sub-tasks, \BFsmall and \BFmedium, depending on the size of the involved methods \cite{Tufano:tosem2019}.

\textbf{Injection of Code Mutants (MG)}.
For the MG task we exploited one of the two datasets provided by Tufano \etal \cite{Tufano:icse2019}: \MGident and \MG{ident-lit}. In both datasets each instance is represented by a triple $\langle  m_\mathit{f}, m_\mathit{b}, M \rangle$, where, similarly to the BF datasets, $m_\mathit{b}$ and $m_\mathit{f}$ are the buggy and fixed version of the snippet, respectively, and $M$ represents the mapping between the abstracted tokens and the code tokens. The first dataset (\MGident) represents the most general (and challenging) case, in which the mutated version, $m_\mathit{b}$, can also contain new tokens (\ie identifiers, types, or method names) not contained in the version provided as input ($m_\mathit{f}$). \MG{ident-lit}, instead, only contains samples in which the mutated version contains a subset of the tokens in the non-mutated code. In other words, \MG{ident-lit} represents a simplified version of the task. For this reason, we decided to focus on the most general scenario and we only use the \MGident dataset. 

\textbf{Generation of Assertions in Test Methods (AG)}.
For the AG task we used the dataset provided by Watson \etal \cite{Watson:icse2020} containing triplets $\langle T, TM_n, A\rangle$, where $T$ is a given test case, $TM_n$ is the \textit{focal} method tested by $T$, \ie the last method called in $T$ before the assert \cite{trace_link}, and $A$ is the assertion that must be generated (output). For such a task, we use two versions of the dataset: \AGraw, which contains the raw source code for the input ($T + TM_n$) and the output ($A$), and \AGabs, which contains the abstracted version of input and output, \ie $src2abs(T + TM_n)$ and $src2abs(A)$, respectively. These are the same datasets used in the original paper.

\textbf{Code Summarization (CS)}.
For code summarization, we exploited the dataset provided by Haque \etal \cite{haque:2020} containing 2,149,120 instances, in which each instance is represented by a tuple $\langle S, A_S, C_S, D \rangle $, where $S$ represents the raw source code of the method, $A_S$ is its AST representation, $C_S$ is the code of other methods in the same file, and $D$ is the summary of the method, \ie the textual description that the model should generate \cite{haque:2020}. For this specific task, we consider a variation of the original dataset to make it more coherent with the performed pre-training. In particular, since in the pre-training we did not use any AST representation of code, we decided to experiment with the T5 model in a more challenging scenario in which only the raw source code to summarize (\ie $S$) is available to the model. Therefore, the instances of our dataset are represented by tuples $\langle S, D \rangle$: We train our model to predict $D$ given only $S$.

%
%

\subsubsection{Decoding Strategy}
Once the models have been trained, different decoding strategies can be used to generate the output token streams. T5 allows to use both \textit{greedy decoding} and \textit{Beam-search}. When generating an output sequence, the greedy decoding selects, at each time step $t$, the symbol having the highest probability. The main limitation of greedy decoding is that it only allows the model to generate one possible output sequence (\eg one possible bug fix) for a given input (\eg the buggy method).

Beam-search is an alternative decoding strategy previously used in many DL applications \cite{DBLP:journals/corr/abs-1211-3711, boulanger2013audio, DBLP:journals/corr/BahdanauCB14, Raychev:2014:CCS:2594291.2594321}. Unlike greedy decoding, which keeps only a single hypothesis during decoding, beam-search of order $K$, with $K > 1$, allows the decoder to keep $K$ hypotheses in parallel: At each time step $t$, beam-search picks the $K$ hypotheses (\ie sequences of tokens up to $t$) with the highest probability, allowing the model to output $K$ possible output sequences.

We used Beam-search to provide several output sequences given a single input, and report results with different $K$ values. It is worth noting that having a large $K$ increases the probability that one of the output sequences is correct, but, on the other hand, it also increases the cost of manually analyzing the output for a user (\ie a developer, in our context).

\subsubsection{Data Balancing for the multi-task model}
The datasets we use for fine-tuning have different sizes, with the one for code summarization dominating the others (see \tabref{tab:finetuning_datasets}). This could result in an unbalanced effectiveness of the model on the different tasks. In our case, the model could become very effective in summarizing code and less in the other three tasks. However, as pointed out by Arivazhagan \etal \cite{DBLP:journals/corr/abs-1907-05019}, there is no free lunch in choosing the balancing strategy when training a multi-task model, with each strategy having its pros and cons (\eg oversampling of less represented datasets negatively impacts the performance of the most representative task). For this reason, we decide to experiment with both strategies. In the first strategy, we follow the true data distribution when creating each batch. In other words, we sample instances from the tasks in such a way that each batch during the training has a proportional number of samples accordingly to the size of the training dataset. For the second strategy, we train a multi-task pre-trained model using a balanced sampling strategy. In other words, we feed the T5 model with batches of data having exactly the same number of samples per task randomly selected during the fine-tuning.

The results we obtained confirm the findings of Arivazhagan \etal \cite{DBLP:journals/corr/abs-1907-05019}. In particular, when using the first training sampling strategy (\ie proportional sampling), the performance of the tasks having a large training dataset (\ie \AGabs, \AGraw, \CS) had a boost. In contrast, when using the second strategy (\ie balanced sampling), the performance increases for those tasks whose training dataset is small with, however, a price to pay for the other three tasks.  
Nonetheless, since the observed differences in performance are not major and each strategy has its pros and cons, we decided to discuss in this paper the results achieved using the proportional sampling schema, as we did in \cite{mastropaolo2021studying}. 

The results of the proportional sampling are available in our replication package \cite{replication}.

\newcommand{\RQ}[1]{RQ\textsubscript{#1}}
\newcommand{\Shared}[1]{$\mathit{Shared}_{\mathit{#1}}$}
\newcommand{\OnlyOurs}[1]{$\mathit{OnlyT5}_{\mathit{#1}}$}
\newcommand{\OnlyBaseline}[1]{$\mathit{OnlyBL}_{\mathit{#1}}$}
\section{Study Design} \label{sec:design}

\begin{table*}[t]
	\caption{Baselines and evaluation metrics for the tasks.}
	\centering
	\begin{tabular}{lllll}
		\toprule
		\textbf{Task}     & \textbf{Baseline}        & \textbf{Accuracy@K}      & \textbf{BLEU-n}         & \textbf{ROUGE LCS} \\
		\midrule
		Automatic Bug-Fixing                & \cite{Tufano:tosem2019}  & $\{1, 5, 10, 25, 50\}$   &   -                      &         -       \\
		Injection of Code Mutants                & \cite{Tufano:icsme2019}  & $\{1\}$                  & $\{A\}$        &       -         \\
		Generation of Asserts in Test                & \cite{Watson:icse2020}   & $\{1, 5, 10, 25, 50\}$   &   -                       &   -             \\
		Code Summarization                & \cite{haque:2020}        &  -                        & $\{1, 2, 3, 4, A\}$                 & $\{P, R, F\}$  \\
		\bottomrule
	\end{tabular}

\label{tab:design}
\end{table*}

We aim at investigating the performance of the T5 model on four code-related tasks: \emph{Automatic bug-fixing, Injection of code mutants, Generation of Asserts in Tests and Code Summarization}. The focus of our evaluation is on (i) investigating the extent to which \emph{transfer learning} is beneficial when dealing with code-related tasks, studying the impact on performance of both pre-training and multi-task learning; and (ii) comparing the obtained results with representative state-of-the-art techniques. The \textit{context} is represented by the datasets introduced in \secref{sec:related}, \ie the ones by Tufano \etal for bug fixing \cite{Tufano:tosem2019} and injection of mutants \cite{Tufano:icsme2019}, by Watson \etal for assert statement generation \cite{Watson:icse2020}, and by Haque \etal for code summarization \cite{haque:2020}. We aim at answering the following research questions (RQs):

\begin{itemize}

\item \textbf{RQ$_{1}$:}\textit{What are the performances of the T5 model when supporting code-related tasks?}
With RQ$_{1}$ we aim at understanding the extent to which T5 can be used to automate code-related tasks, investigating the performance achieved by the model on the four experimented tasks. In the context of RQ$_{1}$, we also investigate the impact of transfer learning on performance:

\begin{itemize}

\item \textbf{RQ$_{1.1}$:}\textit{ What is the role of pre-training on the performances of the T5 model for the experimented code-related tasks?}

With RQ$_{1.1}$ we aim at investigating the boost in performance (if any) brought by pre-training the models on a software-specific dataset.

\item \textbf{RQ$_{1.2}$:}\textit{ What is the role of multi-task learning on the performances of the T5 model for the experimented code-related tasks?} 
RQ$_{1.2}$ analyzes the influence of the \emph{multi-task learning} (\ie training a single model for all four tasks) on the model's performance.

\end{itemize}

\item \textbf{RQ$_{2}$:}\textit { What are the performances of T5 as compared with state-of-the-art baselines?} 
In RQ$_{2}$ we compare the performances achieved by the T5 model against the ones achieved by the baseline approaches. In this regard, we run T5 on the same test sets used in the four original papers presenting automated solutions for the code-related tasks we target.

\end{itemize}

\subsection{Data Collection and Analysis}
\label{sub:dataCollection}

As explained in \secref{subsec:finetuning}, we experimented with different variants of the T5: (i) \emph{no pre-training} (\ie four models each fine-tuned for one of the supported tasks, without any pre-training); (ii) \emph{pre-training single task} (\ie four models each fine-tuned for one of the supported tasks, with  pre-training); and (iii) \emph{pre-training multi-task} (\ie one model pre-trained and fine-tuned for all four tasks). These nine models have all been run on the test sets made available in the works presenting our four baselines and summarized in \tabref{tab:finetuning_datasets}. Once obtained the predictions of the T5 models on the test sets related to the four tasks, we compute the evaluation metrics reported in \tabref{tab:design}. We use different metrics for the different tasks, depending on the metrics reported in the papers that introduced our baselines.

\textbf{Accuracy@K} measures the percentage of cases (\ie instances in the test set) in which the sequence predicted by the model equals the oracle sequence (\ie perfect prediction). Since we use beam-search, we report the results for different $K$ values (\ie 1, 5, 10, 25, and 50), as done in \cite{Tufano:tosem2019} (BF) and \cite{Watson:icse2020} (AG). Tufano \etal \cite{Tufano:icse2019} do not report results for $K > 1$ for the MG task. Thus, we only compute $K = 1$.

\textbf{BLEU score} (Bilingual Evaluation Understudy) \cite{Papineni:2002} measures how similar the candidate (predicted) and reference (oracle) texts are. Given a size $n$, the candidate and reference texts are broken into \textit{n}-grams and the algorithm determines how many \textit{n}-grams of the candidate text appear in the reference text. The BLEU score ranges between 0 (the sequences are completely different) and 1 (the sequences are identical).
We use different BLEU-\textit{n} scores, depending on the ones used in the reference paper of the baseline (see \tabref{tab:design}). For the CS task, we report BLEU-\{1, 2, 3, 4\} and their geometric mean (\ie BLEU-A); for the MG task we only report BLEU-A.

\textbf{ROUGE} (Recall-Oriented Understudy for Gisting Evaluation) is a set of metrics for evaluating both automatic summarization of texts and machine translation techniques in general \cite{lin2004rouge}. ROUGE metrics compare an automatically generated summary or translation with a set of reference summaries (typically, human-produced). We use the ROUGE LCS metrics based on the Longest Common Subsequence for the CS task \cite{haque:2020}. Given two token sequences, $X$ and $Y$, and their respective length, $m$ and $n$, it is possible to compute three ROUGE LCS metrics: $R$ (recall), computed as $\frac{LCS(X, Y)}{m}$, $P$ (precision), computed as $\frac{LCS(X, Y)}{n}$, and F (F-measure), computed as the harmonic mean of $P$ and $R$.\smallskip


The computed metrics are used to select what the best training strategy for the T5 is (\ie \emph{no pre-training}, \emph{pre-training single task}, or \emph{pre-training multi-task}). We also statistically compare the performance of these three strategies for each task using the McNemar's test \cite{mcnemar1947note}, which is a proportion test suitable to pairwise compare dichotomous results of two different treatments. We statistically compare each pair of training strategy in our study (\ie \emph{no pre-training} \emph{vs} \emph{pre-training single task}, \emph{no pre-training} \emph{vs} \emph{pre-training multi-task}, \emph{pre-training single task} \emph{vs} \emph{pre-training multi-task}) in terms of their Accuracy@1 (\ie perfect predictions) for each of the four experimented tasks. To compute the test results for two training strategies $T_1$ and $T_2$, we create a confusion matrix counting the number of cases in which (i) both $T_1$ and $T_2$ provide a correct prediction, (ii) only $T_1$ provides a correct prediction, (iii) only $T_2$ provides a correct prediction, and (iv) neither $T_1$ nor $T_2$ provide a correct prediction. We complement the McNemar's test with the Odds Ratio (OR) effect size. Also, since we performed multiple comparisons, we adjusted the obtained $p$-values using the Holm's correction \cite{Holm1979a}.

The best model output of this analysis has then been used to compare the best T5 model with the four baselines by using the performance metrics reported in \tabref{tab:design}. Moreover, we also statistically compare the Accuracy@1 of the T5 and of the baselines using the same procedure previously described (\ie McNemar's test with the OR effect size). We also perform a complementarity analysis: We define the sets of perfect predictions generated by the T5 ($\mathit{PP}_{\mathit{T5}_d}$) and by the baseline ($\mathit{PP}_{\mathit{BL}_d}$) with a beam size $K=1$. Then, for each task and dataset we compute three metrics:

\footnotesize
$$
\mathit{Shared}_{\mathit{d}} = \frac{|\mathit{PP}_{\mathit{T5}_d} \cap \mathit{PP}_{\mathit{BL}_d}|}{|\mathit{PP}_{\mathit{T5}_d} \cup \mathit{PP}_{\mathit{BL}_d}|}
$$

\noindent\begin{minipage}{.5\linewidth}
	$$
	\mathit{OnlyT5}_{\mathit{d}} = \frac{|\mathit{PP}_{\mathit{T5}_d} \setminus \mathit{PP}_{\mathit{BL}_d}|}{|\mathit{PP}_{\mathit{T5}_d} \cup \mathit{PP}_{\mathit{BL}_d}|}
	$$
\end{minipage}%
\begin{minipage}{.5\linewidth}
	$$
	\mathit{OnlyBL}_{\mathit{d}} = \frac{|\mathit{PP}_{\mathit{BL}_d} \setminus \mathit{PP}_{\mathit{T5}_d}|}{|\mathit{PP}_{\mathit{T5}_d} \cup \mathit{PP}_{\mathit{BL}_d}|}
	$$
	\smallskip

\end{minipage}
\normalsize

\smallskip

\Shared{d} measures the percentage of perfect predictions shared between the two compared approaches on the dataset $d$, while \OnlyOurs{d} and \OnlyBaseline{d} measure the percentage of cases in which the perfect prediction is only generated by T5 or the baseline, respectively, on the dataset $d$.

We also present an ``inference time'' analysis: we compute the time needed to run T5 on a given input. We run such an experiment on a laptop equipped with a 2.3GHz 8-core 9th-generation Intel Core i9 and 16 GB of RAM, \rev{using the CPU to run the DL model}. We do this for different beam search sizes, with $K \in \{1, 5, 10, 25, 50\}$. For each $K$, we report the average inference time (in seconds) on all the instances of each task. \rev{Besides that, we also report the training time (in hours) for the nine different models involved in our study, \ie \emph{no pre-training} (four models, one for each task), \emph{pre-training single task} (+4 models), and \emph{pre-training multi-task} (one model pre-trained and fine-tuned for all four tasks). For the training we used a 2x2 TPU topology (8 cores) from Google Colab with a batch size of 128, with a sequence length (for both inputs and targets) of 512 tokens.}

Finally, we discuss qualitative examples of predictions generated by T5 and by the baselines to give a better idea to the reader about the capabilities of these models in supporting the four code-related tasks.

\subsection{Hyperparameter Tuning} \label{hp-tuning}
Before running the T5 models on the test sets, we performed a hyperparameter tuning on the evaluation sets from \tabref{tab:finetuning_datasets}, to decide the best configuration to run. This was done for all nine models we built (\eg with/without pre-training, with/without multi-task learning). 

For the \textit{pre-training} phase, we use the default parameters defined for the T5 model \cite{raffel2019exploring}. Such a phase, indeed, is task-agnostic, and hyperparameter tuning would provide limited benefits. Instead, we tried different learning rate strategies for the \textit{fine-tuning} phase. Especially, we tested four different learning rates: (i) \textit{Constant Learning Rate} (C-LR): the learning rate is fixed during the whole training; (ii) \textit{Inverse Square Root Learning Rate} (ISR-LR): the learning rate decays as the inverse square root of the training step; (iii) \textit{Slanted Triangular Learning Rate \cite{howard2018universal}} (ST-LR): the learning rate first linearly increases and then linearly decays to the starting learning rate; (iv) \textit{Polynomial Decay Learning Rate} (PD-LR): the learning rate decays polynomially from an initial value to an ending value in the given decay steps. \tabref{tab:hyperparameter:types} reports the specific parameters we use for each scheduling strategy. 

\begin{table}[h]
	\vspace{0.4cm}
	\centering
		\caption{Learning-rates tested for hyperparameter tuning.}
	\begin{tabular}{ll}
		\toprule
		\textbf{Learning Rate Type}   & \textbf{Parameters}\\
		\midrule
		Constant            & $\mathit{LR} = 0.001$        \\
		\midrule
		Inverse Square Root 
		& $\mathit{LR}_{\mathit{starting}} = 0.01$ \\
		& $\mathit{Warmup} = 10,000$ \\
		\midrule
		Slanted Triangular  
		& $\mathit{LR}_{\mathit{starting}} = 0.001$ \\
		& $\mathit{LR_{\mathit{max}}} = 0.01$\\
		& $\mathit{Ratio} = 32$\\
		& $\mathit{Cut} = 0.1$\\
		\midrule
		Polynomial Decay    
		& $\mathit{LR}_{\mathit{starting}} = 0.01$\\
		& $\mathit{LR}_{\mathit{end}} = 0.001$\\
		& $\mathit{Power} = 0.5$\\
		\bottomrule
	\end{tabular}
   \vspace{0.2cm}
	\label{tab:hyperparameter:types}
\end{table}

In total, we fine-tuned 36 models (\ie nine models with four different schedulers) for 100k steps each. To select the best configuration for each training strategy, we compute the following metrics: for BF and AG, we compute the percentage of perfect predictions achieved on the evaluation set with the greedy decoding strategy (Accuracy@1); for MG, we compute the BLEU score \cite{Papineni:2002}; for CS, we compute BLEU-A, the geometric average of the BLEU-\{1,2,3,4\} scores \cite{Papineni:2002}. Basically, for each task we adopt one of the evaluation metrics used in the original paper. The complete results of the hyperparameters tuning phase are reported in our replication package \cite{replication}.


\begin{table*}[t]
	\centering
			\caption{Overall results achieved by the T5 model for each tasks. The best configuration is highlighted in bold}
		\resizebox{\textwidth}{!}{

		\label{tab:t5-results}
	
		\begin{tabular}{lllrrrrrrrrr}
			\hline
			\vspace{0.03cm}
			\textbf{Task} & \textbf{Dataset} & \textbf{Model Configuration} & \bf{Accuracy@1}
			& \bf{Accuracy@5} & \bf{Accuracy@10} &  \bf{Accuracy@25}&  \bf{Accuracy@50} & \bf{BLEU-A} \\
			\hline
			
			\multirow{6}{*}{Automatic Bug-Fixing} &\multirow{3}{*}{\BFsmall} & \emph{no pre-training} & \textbf{16.70\%}  & 29.88\%  & 34.37\%  & 39.57\%  & 42.86\% & -\\
			&&\emph{pre-training single task} & 15.08\% & 32.08\% & 37.01\%  & 42.51\%  & 45.94\%  & -\\
			&& \emph{pre-training multi-task} & 11.61\% & \textbf{35.64\%}  & \textbf{43.87\%}  & \textbf{52.88\%} & \textbf{57.70\%} & - \\\cline{2-9}
			
			 &\multirow{3}{*}{\BFmedium} & \emph{no pre-training} & 10.50\% & 17.60\%  & 20.53\%  & 24.38\%  & 27.62\% & -\\
			&&\emph{pre-training single task} & \textbf{11.85\%}  & \textbf{19.41\%} & 23.28\%  & 28.60\%  & 32.43\%  & -\\
			&& \emph{pre-training multi-task} & 3.65\% & 19.17\%  & \textbf{24.66\%}  & \textbf{30.52\%} & \textbf{35.56\%} & -\\ \hline
			
			\multirow{3}{*}{Injection of Code Mutants} &\multirow{3}{*}{\MGident} & \emph{no pre-training} & 25.78\%  & -  & -  & -  & - & 78.26\%\\
			&&\emph{pre-training single task} & 28.72\%  & - & -  & -  & -  & \textbf{78.69\%}\\
			&& \emph{pre-training multi-task} & \textbf{28.92\%} & -  & -  & - & - & 78.29\%\\ \hline
			
			\multirow{6}{*}{Generation of Asserts in Test} &\multirow{3}{*}{\AGraw} & \emph{no pre-training} & 60.95\%  & 59.14\%  & 62.41\%  & 69.05\%  & 71.97\% & -\\
			&&\emph{pre-training single task} & \textbf{68.93\%}  & \textbf{75.95\%} & \textbf{77.70\%}  & \textbf{79.24\%}  & \textbf{80.22\%}  & -\\
			&& \emph{pre-training multi-task} & 58.60\%& 66.90\%  & 70.31\%  & 73.19\% & 74.58\% & -\\\cline{2-9}
			
			  &\multirow{3}{*}{\AGabs} & \emph{no pre-training} & 47.81\%  & 49.60\%  & 55.04\%  & 64.28\%  & 68.57\% & -\\
			&&\emph{pre-training single task} & \textbf{56.11\%}  & \textbf{71.26\%} & \textbf{74.32\%}  & \textbf{76.67\%}  & \textbf{78.02\%}  & -\\
			&& \emph{pre-training multi-task} & 44.90\%& 63.40\%  & 68.23\%  & 73.04\% & 73.12\% & -\\ \hline
			
			\multirow{3}{*}{Code Summarization} &\multirow{3}{*}{\CS} & \emph{no pre-training} & 11.80\%  & -  & -  & -  & - & 24.67\%\\
			&&\emph{pre-training single task} & \textbf{12.02\%}  & - & -  & -  & -  & \textbf{25.21\%}\\
			&& \emph{pre-training multi-task} &11.45\% & -  & -  & - & - & 24.90\%\\ \hline
		
			\hline
		\end{tabular}

}
	\vspace{0.1cm}

\end{table*}

\newcommand{\about}{$\sim$}
\section{Results Discussion} \label{sec:results}

We discuss our results accordingly to the formulated RQs.

\subsection{Performance of T5 (RQ$_{1}$) and impact of transfer learning on performance (RQ$_{1.1}$-RQ$_{1.2}$)}

\tabref{tab:t5-results} reports the performance achieved by the different variants of the T5 model that we experimented with. For each task (\eg Automatic Bug-Fixing) and for each dataset (\eg BFsmall), performance metrics are reported for the three adopted training strategies (\ie no pre-training, pre-training single task, and pre-training multi-task). For readability reasons, we only report the BLEU-A, but the results of the other BLEU scores (\eg BLEU-4) are available in our online appendix \cite{replication}. \rev{The pre-training multi-task setting is the same as used in our ICSE'21 paper \cite{mastropaolo2021studying} that this work extends. Note that for some tasks (\eg \AGraw) the results reported in Table 5 are different as compared to the ones reported in the ICSE paper. This is due to two changes we performed in our experimental pipeline. First, as compared to the ICSE paper, we updated our scripts to exploit the latest T5 version available as of today (\ie T5 0.9.2 - \url{https://libraries.io/pypi/t5/0.9.2}) and re-executed all of our experiments. Second, in our ICSE paper we lower-cased the source code before providing it as input to the T5. However, we realized that when working with Java raw code (see \eg the \AGraw task), it is important to keep such information considering the wide adoption of the camelCase naming convention in such a language.}

\tabref{tab:mc-test} reports the results of the statistical analysis we performed using the McNemar's test \cite{mcnemar1947note} to identify (if any) statistical differences in terms of Accuracy@1 when using different training strategies.

\begin{table*}[h!]
	\centering
	\caption{McNemar's test (adj. $p$-value and OR) considering only accuracy@1 matches as correct predictions}

		\label{tab:mc-test}

		\begin{tabular}{lllrrrrrr}
			\hline
			\vspace{0.03cm}
			\textbf{Task} & \textbf{Dataset} & \textbf{Model Configuration} & \textbf{\emph{p}-value} & \textbf{OR} & \\
			\hline
			
			\multirow{6}{*}{Automatic Bug-Fixing} &\multirow{3}{*}{\BFsmall} & \emph{no pre-training} \emph{vs} \emph{pre-training single task} &$<0.001$  & 0.77  \\
			&&\emph{no pre-training} \emph{vs} \emph{pre-training multi-task} & $<0.001$ & 0.46 \\
			&& \emph{pre-training multi-task} \emph{vs} \emph{pre-training single task} & $<0.001$ & 1.67   \\\cline{2-5}
			
			&\multirow{3}{*}{\BFmedium} & \emph{no pre-training} \emph{vs} \emph{pre-training single task} & $<0.001$  & 1.56  \\
			&&\emph{no pre-training} \emph{vs} \emph{pre-training multi-task}& $<0.001$ & 0.12 \\
			&& \emph{pre-training multi-task} \emph{vs} \emph{pre-training single task} & $<0.001$ & 8.56   \\

			\midrule
			
			\multirow{3}{*}{Injection of Code Mutants} &\multirow{3}{*}{\MGident} & \emph{no pre-training} \emph{vs} \emph{pre-training single task} & $<0.001$ & 1.51  \\
			&&\emph{no pre-training} \emph{vs} \emph{pre-training multi-task} & $<0.001$ & 1.38 \\
			&& \emph{pre-training multi-task} \emph{vs} \emph{pre-training single task} & 0.75 & 0.99   \\
			
			\midrule
			
			\multirow{6}{*}{Generation of Asserts in Test} &\multirow{3}{*}{\AGraw} & \emph{no pre-training} \emph{vs} \emph{pre-training single task} & $<0.001$  & 3.39  \\
			&&\emph{no pre-training} \emph{vs} \emph{pre-training multi-task} & $<0.001$ & 0.71 \\
			&& \emph{pre-training multi-task} \emph{vs} \emph{pre-training single task} & $<0.001$ & 4.95   \\\cline{2-5}
			
			&\multirow{3}{*}{\AGabs}& \emph{no pre-training} \emph{vs} \emph{pre-training single task} & $<0.001$  & 2.55  \\
			&&\emph{no pre-training} \emph{vs} \emph{pre-training multi-task} & $<0.001$ & 0.74 \\
			&& \emph{pre-training multi-task} \emph{vs} \emph{pre-training single task} & $<0.001$ & 2.93   \\
			
			\midrule
			
			\multirow{3}{*}{Code Summarization} &\multirow{3}{*}{\CS} & \emph{no pre-training} \emph{vs} \emph{pre-training single task} & $<0.001$  & 1.13  \\
			&&\emph{no pre-training} \emph{vs} \emph{pre-training multi-task} & $<0.001$ & 0.83 \\
			&& \emph{pre-training multi-task} \emph{vs} \emph{pre-training single task} & $<0.001$ & 1.40   \\
			
			\hline
		\end{tabular}

\scriptsize

\end{table*} 

Focusing on the Accuracy@1, it is evident that there is no training strategy being the best one across all tasks and datasets. In particular: \emph{no pre-training} works better on the \BFsmall dataset for automatic bug-fixing; \emph{pre-training single task} works better on the \BFmedium dataset for automatic bug-fixing, on both datasets related to the generation fo assert statements, and for the code summarization task; finally, \emph{pre-training multi-task} works better for the injection of code mutants. Overall, the \emph{pre-training single task} strategy seems to be the best performing strategy. Indeed, even when it is not the first choice for a given task/dataset, it is the second best-performing training strategy. Also, by looking at \tabref{tab:mc-test} we can observe that:

\begin{enumerate}
\item When \emph{pre-training single task} is the best strategy, its performance in terms of Accuracy@1 are significantly better ($p$-value $<$ 0.001) than the second best-performing strategy, with ORs going from 1.13 (for \CS) to 3.39 (\AGraw). This means that chances of getting a perfect predictions using this strategy are 13\% to 339\% higher when using this strategy as compared to the second choice.

\item  When \emph{pre-training single task} is not the best strategy, but the second choice, the difference in Accuracy@1 is not significant when compared to \emph{pre-training multi-task} for \MGident. The only significant difference is the one in favor of \emph{no pre-training} on \BFsmall, with an OR of 0.77.
\end{enumerate}

For these reasons, in our RQ$_{2}$ we will compare the T5 using the \emph{pre-training single task} strategy against the baselines.

A few observations can be made based on the findings in \tabref{tab:t5-results}. First, the additional pre-training is, as expected, beneficial. Indeed, on five out of the six datasets the T5 performs better with pre-training. Second, the multi-task setting did not help in most of cases. Indeed, with the exception of \MGident in which the performance of \emph{pre-training single task} and \emph{pre-training multi-task} are basically the same, the \emph{single task} setting performs always better. Such a result, while surprising at a first sight, can be explained by diverse types of input/output handled by the models across the four tasks. Indeed, (i) the datasets related to automatic bug-fixing and \AGabs include abstracted code instances as input/output; (ii) the dataset used for code mutants and \AGraw feature raw code instances as input/output; and (iii) the one for code summarization has raw source code as input and natural language text as output. Basically, given the different formats, the transfer learning across different tasks is likely to hinder the model rather than helping it. 

Differently, the pre-training dataset features all three input/output representations and, thus, provides the model with a basic knowledge about all of them that, as a result, boosts performance. 

While we will discuss more in depth the performance of the T5 model when comparing it to the considered baselines (\secref{sec:t5-vs-baselines}), it is also worth commenting on the ability of the T5 to generate correct predictions, namely outputs that are identical to the reference ones (\eg a method summary equal to the one manually written by developers). Quite impressive are the performances achieved on the generation of assert statements, especially on the dataset dealing with raw source code, in which the T5 correctly predicts 68.93\% of assert statements with a single guess (75.95\% when using five guesses). The Accuracy@1 is instead much lower for the other tasks, ranging between 11.85\% (fixing bugs in the most challenging \BFmedium dataset) up to 28.72\% when injecting mutants. Also worth noticing is the 12.02\% of code summaries generated by the T5 that are identical to the manually written ones. In the next subsection, together with a comparison of our model with the baselines, we present qualitative examples of predictions generated by the T5.

\begin{table*}[h!]
	\centering
	\caption{Top-20 AST operations needed to fix bugs in our dataset (see ``Oracle'' column) and their presence in correct predictions generated by T5 and the baseline}
	\scriptsize
	\label{tab:astBugs}
	\begin{tabular}{lrrrrrrr}
		\toprule
		\multicolumn{8}{c}{{\bf Delete}}\\\midrule
		& \multicolumn{3}{c}{\BFsmall} && \multicolumn{3}{c}{{\BFmedium}}\\ \cmidrule{2-4} \cmidrule{6-8}
		
		& {\bf Oracle} & {\bf Baseline \cite{Tufano:tosem2019}} & {\bf T5} 
		&& {\bf Oracle}  &  {\bf Baseline \cite{Tufano:tosem2019}} &  {\bf T5}\\\midrule
		
		Delete TypeAccess at Invocation& 2,016 & 402 & 450 && 1,926 & 125 & 250\\
		Delete Invocation at Block & 1,444 & 294 & 326 && 1,315 & 159 & 240\\
		Delete TypeAccess at ThisAccess & 821 & 92 & 134 && 598 & 32 & 81\\
		Delete VariableRead at Invocation & 818 & 51 &106 && 1,106 & 61 & 126\\
		Delete FieldRead at BinaryOperator & 479 & 92 & 100 && 651 & 66 & 116\\\midrule

		\multicolumn{8}{c}{{\bf Insert}}\\\midrule
		& \multicolumn{3}{c}{\BFsmall} && \multicolumn{3}{c}{{\BFmedium}}\\ \cmidrule{2-4} \cmidrule{6-8}
		
		& {\bf Oracle} & {\bf Baseline \cite{Tufano:tosem2019}} & {\bf T5} 
		&& {\bf Oracle}  &  {\bf Baseline \cite{Tufano:tosem2019}} &  {\bf T5}\\\midrule
		
		Insert Block at If & 486 & 3 & 28 && 828 & 3 & 48\\
		Insert Literal at BinaryOperator & 468 & 5 & 27 && 736 & 0 & 37\\
		Insert If at Block & 406 & 2 &22 && 659 & 0 & 33\\
		Insert BinaryOperator at If & 380 & 3 & 23 && 634 & 0 & 36\\
		Insert VariableRead at Invocation & 328 & 10 & 33 && 675 & 0 & 38\\\midrule

		\multicolumn{8}{c}{{\bf Move}}\\\midrule
		& \multicolumn{3}{c}{\BFsmall} && \multicolumn{3}{c}{{\BFmedium}}\\ \cmidrule{2-4} \cmidrule{6-8}
		
		& {\bf Oracle} & {\bf Baseline \cite{Tufano:tosem2019}} & {\bf T5} 
		&& {\bf Oracle}  &  {\bf Baseline \cite{Tufano:tosem2019}} &  {\bf T5}\\\midrule
		
		Move Invocation from Block to Invocation & 633 & 17 & 61 && 1,005 & 4 & 86\\
		Move VariableRead from Invocation to VariableRead & 158 & 7 & 11&& 281 & 2 & 19\\
		Move Assignment from Block to Assignment & 120 & 0 & 13 && 209 & 1 & 19\\
		Move Invocation from BinaryOperator to Invocation & 95 & 7 & 11 && 183 & 1 & 14\\
		Move BinaryOperator from BinaryOperator to BinaryOperator & 68 & 0 & 2 && 174 & 0 & 9\\\midrule
		
		\multicolumn{8}{c}{{\bf Update}}\\\midrule
		& \multicolumn{3}{c}{\BFsmall} && \multicolumn{3}{c}{{\BFmedium}}\\ \cmidrule{2-4} \cmidrule{6-8}
		
		& {\bf Oracle} & {\bf Baseline \cite{Tufano:tosem2019}} & {\bf T5} 
		&& {\bf Oracle}  &  {\bf Baseline \cite{Tufano:tosem2019}} &  {\bf T5}\\\midrule
		
		Update Wra at Method & 280 & 15 & 37 && 191 & 1 & 22\\
		Update TypeAccess at Invocation & 201 & 17 & 41 && 404 & 18 & 115\\
		Update Invocation at Block & 115 & 0 & 8 && 153 & 2& 21\\
		Update VariableRead at Invocation & 101 & 1 &12 && 226 & 0 & 19\\
		Update BinaryOperator at If & 56 & 3 & 8 && 148 & 1 & 12\\\midrule
		
	\end{tabular} 
\end{table*}

\subsection{Competitiveness of the T5 model compared to the baselines ($RQ_2$) }
\label{sec:t5-vs-baselines}

We compare the results achieved by the T5 model when using the \emph{pre-training single task} strategy with the baseline we consider for each task (\tabref{tab:design}). The comparison is depicted in \figref{fig:results}, while \tabref{tab:mc-baselines} shows the results of the statistical tests, and \tabref{tab:overlap} shows the overlap metrics described in \secref{sub:dataCollection}.

\subsubsection{Automatic Bug Fixing (BF)}

When using T5 for automatically fixing bugs, the accuracy achieved using a greedy decoding strategy ($K = 1$) 
differs according to the dataset we consider. For example, the T5 model achieves 15\% of perfect predictions on the \BFsmall dataset against 9\% achieved by the baseline, with an improvement of 6 percentage points, while in the most challenging scenario, (\ie  \BFmedium) our model obtains an improvement of 8 percentage points over the baseline (11\% \emph{vs} 3\%). Such improvements are statistically significant (\tabref{tab:mc-baselines}) with ORs of 2.39 (\BFsmall) and 6.88 (\BFmedium), indicating higher chance of observing a perfect prediction when using the T5 as compared to the baseline. Worth noticing is that as the beam width increases, the performance of the T5 and of the baseline gets closer, with the baseline performing better for $K = 25$ and $K = 50$ on \BFsmall.

Looking at the overlap metrics (\tabref{tab:overlap}), 25.90\% of perfect predictions on \BFsmall and 28.78\% on \BFmedium are shared by the two techniques. The remaining are perfect predictions only with T5 (53.20\% on \BFsmall and 36\% on \BFmedium) or only with the baseline (20.90\% on \BFsmall and 35.16\% on \BFmedium). This indicates that the two approaches are complementary for the bug fixing task suggesting that further improvements could be possible by exploiting customized ML-based bug-fixing techniques.
\rev{To further look into this finding, we analyzed the type of ``code transformation'' that T5 and the baseline were able to learn. With ``code transformation'' we refer to Abstract Syntax Tree (AST) operations needed to correctly transform the input code into the target prediction (\ie the AST operations performed by developers to transform the buggy code into the fixed code). In particular, we used the Gumtree Spoon AST Diff \cite{gumtree} to collect the \emph{Delete}, \emph{Insert}, \emph{Move} and \emph{Update} operations performed on the AST nodes when fixing bugs. Then, for each of these operations, we extracted the 5 most popular ones (\eg the five most popular \emph{Delete} node operations). These 20 AST-level operations (4 types of operations $\times$ 5 most popular for each type)  characterize the successful fixing of bugs/injection of code mutants in the three datasets. The column ``Oracle'' of (\tabref{tab:astBugs}) reports such numbers.  Then, we took the correct predictions generated by T5 and by the baselines and checked the extent to which those predictions feature the ``popular'' AST operations that, accordingly to our oracles, are needed to properly fix bugs. \tabref{tab:astBugs} reports for both techniques and both datasets (\BFsmall and \BFmedium) the number of times the different AST operations were performed by the models.}

\rev{Given the previously discussed superior performance of T5, it is expected to see that it managed to correctly perform the needed AST operations more often than the baseline. However, what is interesting is that there are specific types of operations that are not learned by the baseline while they are successfully implemented by T5. This is especially true for less popular operations such as the \emph{Insert} ones, that require to synthesize new nodes that were not present in the input AST. In \BFmedium, four of the top-five AST Insert operations are never applied by the baseline (see \tabref{tab:astBugs}). Similar results are also obtained for the \emph{Update} operations, while both models work similarly well when the bug-fix mostly requires the deletion of existing AST nodes.}


%

\begin{table}
	\centering
	\caption{McNemer's test considering the correct predictions achieved by the T5 model and the baselines when both techniques generate only one prediction (\ie \emph{accuracy@1})}
	\label{tab:mc-baselines}
	\resizebox{\linewidth}{!}{
		\begin{tabular}{l l r r }
			\toprule
			\emph{Task}&  \emph{Dataset} ($d$) & \textbf{\emph{p}-value} & \textbf{OR} \\
			\midrule
			
			\multirow{2}{*}{Automatic Bug-Fixing} &  \BFsmall             & $<0.001$    & 2.39            \\
			& \BFmedium            & $<0.001$    & 6.88              \\
			
			\midrule
			Injection of Code Mutants & \MGident             & $<0.001$    & 2.95        \\		
			\midrule
			
			\multirow{2}{*}{Generation of Asserts in Test} & 
			\AGabs               &  $<0.001$    & 6.19         \\
			& \AGraw               & $<0.001$    & 43.12               \\
			\midrule
			
			Code Summarization & \CS                  & $<0.001$    &35.56             \\
			\bottomrule
			
		\end{tabular}
	}
\end{table}

\subsubsection{Injection of Code Mutants (MG)}

\begin{table}[h!]
	\centering
	\caption{Top-20 AST operations needed to inject mutants in our dataset (see ``Oracle'' column) and their presence in correct predictions generated by T5 and the baseline}
	\scriptsize
	\label{tab:astMutants}
	
	\resizebox{\linewidth}{!}{
		\begin{tabular}{lrrrrrrr}
			\toprule
			\multicolumn{4}{c}{{\bf Delete}}\\\midrule
			& \multicolumn{3}{c}{\MGident}\\ \cmidrule{2-4} 
			
			& {\bf Oracle} & {\bf Baseline \cite{Tufano:icsme2019}} & {\bf T5}\\\midrule
			
			Delete TypeAccess at Invocation & 387 & 1 & 30 \\
			Delete Return at Block & 327 & 20 & 64 \\
			Delete FieldRead at BinaryOperator & 283 & 0 & 7\\
			Delete FieldRead at Invocation & 242 & 0 &19\\
			Delete Invocation at Block & 236 & 0 & 15 \\\midrule
			
			\multicolumn{4}{c}{{\bf Insert}}\\\midrule
			& \multicolumn{3}{c}{\MGident}\\ \cmidrule{2-4} 
			
			& {\bf Oracle} & {\bf Baseline \cite{Tufano:icsme2019}} & {\bf T5}\\\midrule
			
			Insert TypeAccess at Invocation & 6,230 & 1,125 & 1,744 \\
			Insert Invocation at Block & 3,979 & 860 & 1,183 \\
			Insert TypeAccess at ThisAccess & 2,219 & 479 & 722\\
			Insert VariableRead at Invocation & 2,061 & 245 &466\\
			Insert Block at If & 1,795 & 485 & 671 \\\midrule

			\multicolumn{4}{c}{{\bf Move}}\\\midrule
			& \multicolumn{3}{c}{\MGident}\\ \cmidrule{2-4} 
			
			& {\bf Oracle} & {\bf Baseline \cite{Tufano:icsme2019}} & {\bf T5}\\\midrule
			
			Move Invocation from Block to Invocation & 1,154 & 225 & 356 \\
			Move Invocation from Return to Invocation & 283 & 55 & 105 \\
			Move Return from Block to Return & 224 & 58 & 100\\
			Move Assignment from Block to Assignment & 190 & 26 &56\\
			Move Invocation from Invocation to Invocation & 129 &1 & 27 \\\midrule

			\multicolumn{4}{c}{{\bf Update}}\\\midrule
			& \multicolumn{3}{c}{\MGident}\\\cmidrule{2-4} 
			
			& {\bf Oracle} & {\bf Baseline \cite{Tufano:icsme2019}} & {\bf T5}\\\midrule
			
			Update TypeAccess at Invocation & 923 & 67 & 220 \\
			Update FieldRead at BinaryOperator & 408 & 14 & 63 \\
			Update Wra at Method & 264 & 1 & 31\\
			Update TypeAccess at ThisAccess & 228 & 10 &73\\
			Update TypeReference at Method & 208 & 0 & 25 \\
			\bottomrule
		\end{tabular}
	}
\end{table}

Looking at \figref{fig:results} we can observe that using T5 to generate mutants allows to obtain more accurate results than the baseline, with the Accuracy@1 improving by 12 percentage points, with 1,336 additional perfect predictions. The average BLEU score also improves by \about 0.02 on top of the very good results already obtained by the baseline (\ie 0.77). Minor improvements in BLEU score can still indicate major advances in the quality of the generated solutions \cite{googleblogentry}. Also in this case differences in terms of Accuracy@1 are statistically significant, with the T5 model being more likely to generate correct solutions ($OR=2.95$) as compared to the baseline approach \cite{Tufano:icsme2019} (\tabref{tab:mc-baselines}).

Differently from the bug-fixing task, for the injection of code mutants the percentage of shared perfect predictions (\tabref{tab:overlap}) is slightly higher (33\%) with, however, T5 being the only one generating 50.52\% of perfect predictions as compared to the 16.48\% generated exclusively by the baseline. 

\rev{Similarly to what has been done in the context of the bug-fixing task, \tabref{tab:astMutants} reports the top-20 AST-level operations needed to correctly inject mutants in our dataset. Note that, differently from what observed for the bug-fixing task, injecting mutants mostly requires the insertion of new AST nodes. The trend that we observe is, as expected, the opposite of what we found for the bug-fixing task because the task is the same but with reversed input/output. Indeed, the baseline seems to correctly predict the most popular \emph{Insert} operations in the AST, while it almost ignores the more rare \emph{Delete} ones. T5 instead, covers all top-20 operations.
}

\subsubsection{Generation of Assertions in Test Methods (AG)}

\begin{figure*}[h!]
	\centering
	\captionsetup{justification=centering,margin=2cm}
	\caption{Performance of the T5 model against the experimented baselines.}
	\includegraphics[width=\linewidth]{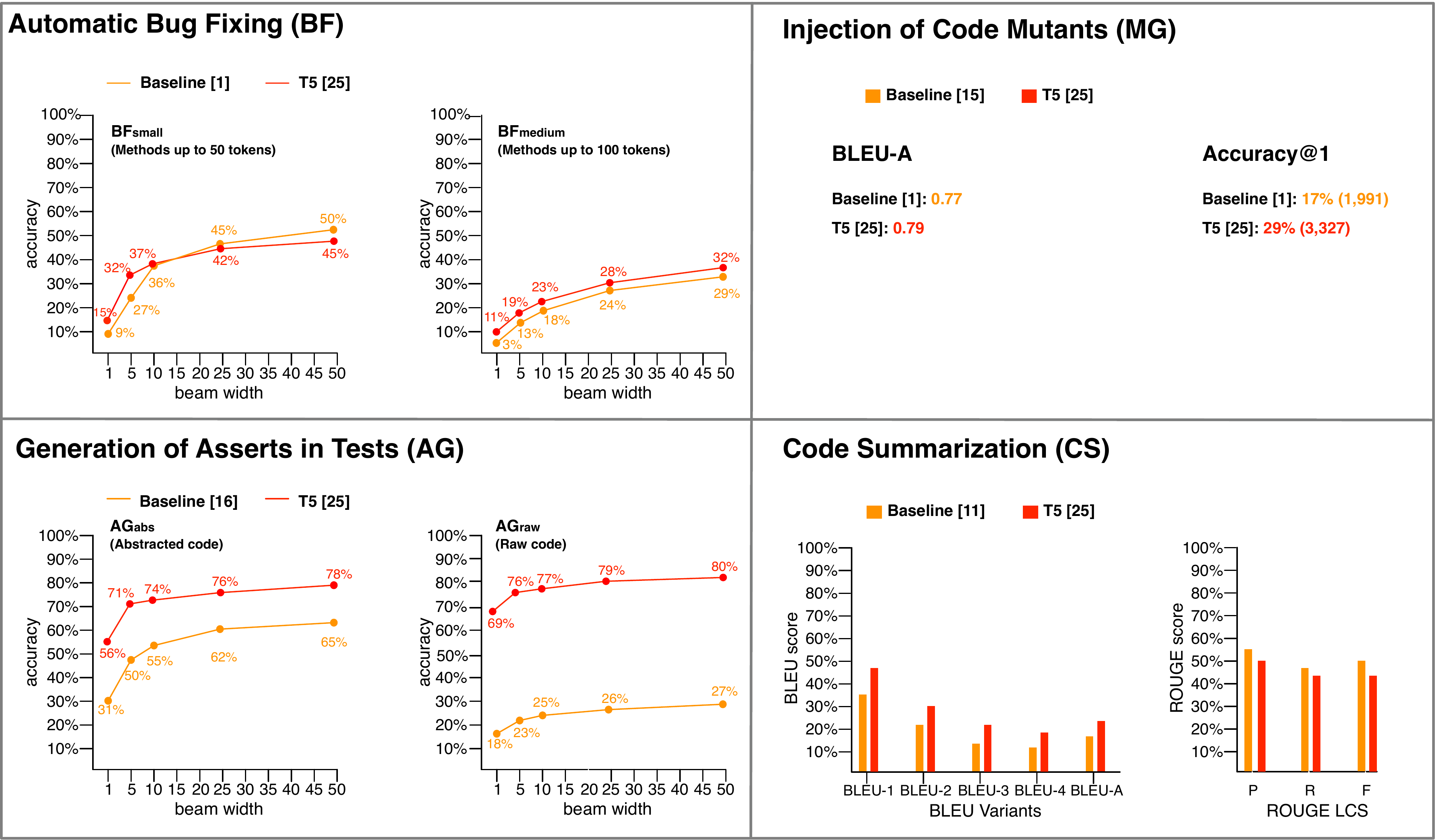}
	\label{fig:results}
\end{figure*}

T5 achieve much better performance in this task as compared to the baseline. The gap is substantial both with (\AGabs) and without (\AGraw) code abstraction (\figref{fig:results}). With abstraction, the T5 achieves a 56\% accuracy at $K=1$ against the 31\% achieved by \emph{ATLAS} \cite{Watson:icse2020}. When both approaches are asked to generate multiple assert statements (\ie $K=5,10,25,50$) the gap in performance ranges between  13-25 percentage points. When using the more challenging non-abstracted dataset \AGraw, T5 achieves even better results. In this regard, when T5 is asked to generate only one assert statement ($K = 1$), the reported accuracy is 51 percentage points higher compared to the baseline , while for larger $K$ values the gap in performance ranges between 51-53 percentage points. The McNemar's test confirms the huge gap in performance between the two techniques, with $ORs$ ranging between 6.19 (\AGabs) and 43.12 (\AGraw).



In terms of overlap, we found a trend similar to the previously discussed task (mutants injection): On \AGabs we have 34.92\% of perfect predictions shared between the two approaches, while the remaining instances are distributed between the ones only predicted by T5 (58.87\%) and the ones only predicted by the baseline (6.21\%). The overlap is much smaller on the \AGraw dataset, with only 9.56\% of the instances correctly predicted by both the approaches, 89.65\% of them correctly predicted only by T5, and 0.79\% only by the baseline.

\subsubsection{Code Summarization (CS)}
On this task, T5 achieves a substantial increase in BLEU score as compared to the baseline. When considering the average BLEU (BLEU-A), the improvement is of $\sim$5 percentage points. On the other hand, it can be noticed that the ROUGE-LCS scores achieved when using T5 are lower than the ones achieved by the baseline ($\sim$5 percentage points lower on the F-measure score). Thus, looking at these metrics, there is no clear winner, but T5 seems to be at least comparable to the baseline. To have something easier to interpret, we compared the two approaches in terms of the number of perfect predictions they generate, despite the fact that such a metric was not used in the original paper \cite{haque:2020}. This means counting the comments generated by a technique that are exactly equal to the ones manually written by humans. T5 managed to generate 12.02\% of perfect predictions (10,929 instances) against the 3.4\% (3,048) of the baseline technique (over 3 $\times$ better). As expected from previous results, the majority of the perfect predictions for this task can be done only using T5 (93.79\%). A limited percentage of perfect predictions is shared (4.79\%), and a minority of instances can be only predicted through the baseline (1.42\%). The McNemar's test highlights a statistical significance in terms of Accuracy@1, with an OR of 35.56.

\begin{table}[h!]
	\centering
	\caption{Overlap metrics for correct predictions generated by the T5 model and the baselines.}
	\label{tab:overlap}
	\resizebox{\linewidth}{!}{
		\begin{tabular}{l l r r r}
			\toprule
			\emph{Task}&  \emph{Dataset} ($d$) & \Shared{d} & \OnlyOurs{d} & \OnlyBaseline{d} \\
			\midrule
			
			\multirow{2}{*}{Automatic Bug-Fixing} &  \BFsmall             & 25.90\%    & 53.20\%      & 20.90\%          \\
			& \BFmedium            & 28.78\%    & 36.06\%      & 35.16\%          \\
			
			\midrule
			Injection of Code Mutants & \MGident             & 33.00\%    & 50.52\%      & 16.48\%          \\		
			\midrule
			
			\multirow{2}{*}{Generation of Asserts in Test} & 
			\AGabs               & 34.92\%    & 58.87\%      & 6.21\%          \\
			& \AGraw               & 9.56\%    & 89.65\%      & 0.79\%           \\
			\midrule
			
			Code Summarization & \CS                  & 4.79\%     & 93.79\%      & 1.42\%           \\
			\bottomrule
			
		\end{tabular}
	}
\end{table}

\subsubsection{Qualitative Analysis}\label{qualitative}
\begin{figure*}[h!]
	\centering
	\captionsetup{justification=centering,margin=2cm}
	\caption{Examples of perfect and alternative predictions}
	\includegraphics[width=\linewidth]{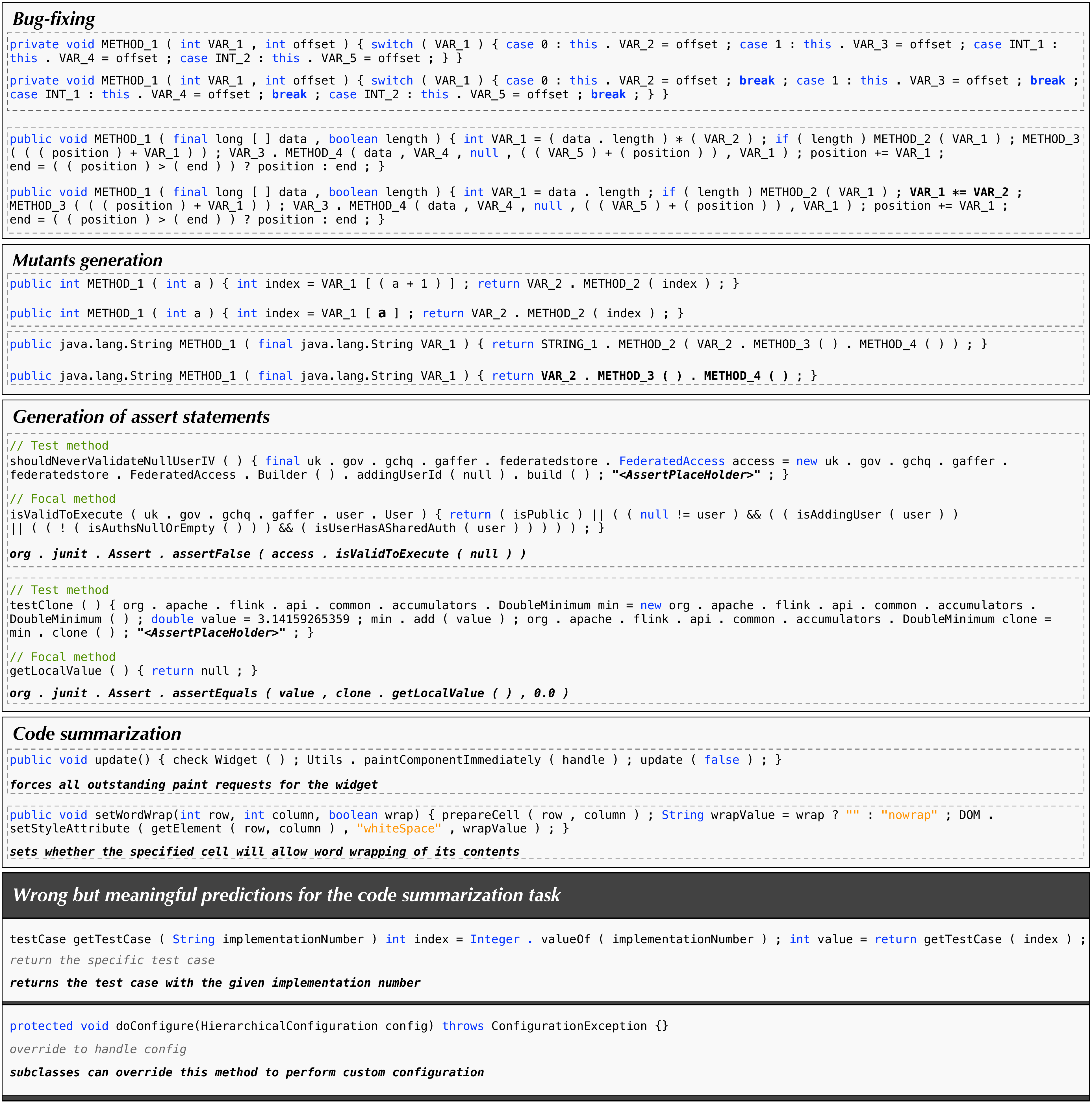}
	\label{fig:qualitative-examples}
\end{figure*}

To give a better idea to the reader about the capabilities of the T5 model in supporting the four code-related tasks, \figref{fig:qualitative-examples} shows two examples of perfect predictions made by T5 for each task. Each example is bordered with a dashed line and shows (i) the input provided by the model, and (ii) the generated output. In particular, in the case of the bug-fixing, mutants injection, and code summarization tasks, the first line shows the input and the second the output. Concerning the generation of assert statements, the first two lines (\ie those marked with ``//Test method'' and ``//Focal method'') represent the input, while the third line shows the generated assert statement. We highlighted in bold the most relevant parts of the output generated by the model. The bottom part of \figref{fig:qualitative-examples} also shows some ``wrong'' predictions (\ie the output of the model is different from the expected target) for the code summarization task, that we will discuss later on.

Concerning the bug-fixing task, in the first example the model adds the \texttt{break} statement to each \texttt{case} of the \texttt{switch} block, thus allowing the program to break out of the \texttt{switch} block after one \texttt{case} block is executed. In the second example, instead, it changes the execution order of a statement as done by developers to fix the bug.

As per the mutants injection, the first example represents an \emph{arithmetic operator deletion}, while the second is a \emph{non void method call mutation} \cite{pitest}. While these transformations might look trivial, it is worth remembering that they are considered as correct since they reproduce real bugs that used to affect these methods. Thus, the model is basically choosing where to mutate and what to mutate in such a way to simulate real bugs (accomplishing one of the main goals of mutation testing).

Both examples of correct prediction we report involve the generation of an assert statement including an invocation to the focal method (\ie the main method tested by the test method). While the first is a rather ``simple'' \texttt{assertFalse} statement, the second required the guessing of the expected value (\ie \texttt{assertEquals}).

Finally, for the code summarization, the two reported examples showcase the ability of T5 to generate meaningful summaries equivalent to the ones manually written by developers. For this task, we also reported in the bottom part of the figure some wrong but still meaningful predictions. In this case, the grey text represent the original summary written by developers, while the bold one has been generated by T5. In both cases, the generated summary is semantically equivalent and even more detailed that the manually written one. 

These two examples help in discussing an important limitation of our analysis: While we assume the correct predictions to be the \emph{only} valuable outputs of T5 and of the experimented baselines, they actually represent a lower-bound for their performance. Indeed, there are other predictions that, even if wrong, could still be valuable for developers, such as the two shown for the code summarization task.

\subsection {Training and Inference Time}

\rev{
\tabref{tab:training-time} reports the training time (in hours) for the nine models we trained. On average, the infrastructure we used for training requires 31.5 seconds every 100 training steps which, given our batch size = 128, means that 12,800 training instances can be processed in 31.5 seconds. Of course, multiple passes (usually referred to as epochs) are needed on the dataset during the training. \tabref{tab:training-time} shows that (i) the pre-training has a cost of $\sim$22h that should be added on top of the fine-tuning cost shown for each task; (ii) as expected, the training time increases with the increase in size of the training dataset, with the \emph{code summarization} task being the most expensive in terms of training time; (iii) clearly, the \emph{multi-task} setting requires to train the model on all tasks, resulting in the highest training time (175h).
}

\begin{table}[h!]
	\caption{Training time (hours) for the trained T5 models}
	\label{tab:training-time}
	\resizebox{\linewidth}{!}{
		\begin{tabular}{lrrrrr}
			\toprule
			\multirow{2}{*}{\textbf{Training}} &  \multirow{2}{*}{\textbf{Bug-fixing}} & \textbf{Mutants} & \textbf{Generation of} & \textbf{Code} & \multirow{2}{*}{\textbf{Multi-Task}}\\
			& & \textbf{generation} & \textbf{assert statements} & \textbf{summarization} & \\\midrule
			No pre-training  &  6.26 & 5.85  & 17.51 & 123.55 & - \\
			Pre-training     &  28.10 & 27.72  & 39.40    & 145.42   & 175.00\\
			\bottomrule
		\end{tabular}
	} 
\end{table}

\rev {\tabref{tab:inference_time} presents, instead,} the results of the inference time analysis (\ie the time needed to run the model on a given input and obtain the prediction). Such analysis allows to understand the extent to which such a model can be used in practice. \tabref{tab:inference_time} reports the inference time in seconds for different $K$ values (\eg with $K=10$ the reported time is the one required by the model to generate 10 possible solutions).

\begin{table}[h!]
	\caption{Inference time with different beam size values.}
	\label{tab:inference_time}
	\resizebox{\linewidth}{!}{
		\begin{tabular}{l l l l l l l}
			\toprule
			$K$ & \BFsmall & \BFmedium & \MGident & \AGabs & \AGraw & \CS  \\
			\midrule                        
			1   & 0.72      & 1.86      & 0.94     & 0.73   & 0.53   & 0.20 \\
			5   & 1.47      & 3.69      & 1.70     & 1.59   & 1.04   & 0.36 \\
			10  & 1.91      & 5.26      & 2.20     & 2.64   & 1.52   & 0.48 \\
			25  & 3.54    & 11.10      & 4.32     & 5.45   & 3.15   & 0.81 \\
			50  & 5.99    & 20.90      & 7.60     & 10.24   & 5.45   & 1.45 \\
			\bottomrule
		\end{tabular}
	}
\end{table}

Concerning the bug-fixing task, the time needed to generate a fix depends on the dataset, since the complexity of the instances they feature is different. In the \BFsmall dataset, the average inference time ranges between 0.72s ($K=1$) and 5.99s ($K=50$), while it is larger on the \BFmedium dataset (1.86s for $K=1$ and 20.90s for $K=50$).
For the injection of code mutants, we observed results comparable to those of \BFsmall: with $K = 1$ the average inference time is 0.94s, while for $K = 50$ it is 7.60s. The generation of assert statement is very fast for low values of $K$ (0.73s for \AGabs and 0.53s for \AGraw with $K = 1$), while it gets slower for higher values of $K$ (10.24 for \AGabs and 5.45 for \AGraw with $K = 50$). Finally, concerning the code summarization task, T5 takes only 0.20s for $K = 1$ and 1.45s for $K = 50$ to output code summaries for a method given as input.

Overall, considering that all the targeted tasks do not have strong real-time constraints (\eg a developer can wait a few seconds for the automated fixing of a bug), the inference times should not hinder the model applicability in practice. \rev{Also, the reported inference times were obtained by running the model on a consumer-level device and by only using CPUs. We also computed the inference time using an Nvidia Tesla P100 GPU equipped with 16GB of VRAM. The achieved results are available in our replication package \cite{replication}. In summary, we observed an average decrease of inference time of $\sim$70\% as compared to the one obtained using the CPU.}

\section{Threats to Validity} \label{sec:threats}

\textbf{Construct validity.} Threats to construct validity concern the relationship between theory and observation. We used existing datasets that are popular and used in the community for both pre-training and fine-tuning our model with minimal additional processing (\eg removal of duplicates after abstraction in the dataset used for the pre-training). Additionally, we have released all of our code and models in our replication study \cite{replication} for verification.

\textbf{Internal validity.} Threats to internal validity concern factors, internal to our study, that could influence its results. Many factors can influence our results, from model architecture, hyperparameter choices, data processing, the data itself, \etc For mitigating these issues, we have adopted methodologies usually employed in DL-based research. Specifically, we performed a detailed analysis of hyperparameter choices as discussed in \secref{hp-tuning}. Concerning the pre-training phase, we used the default T5 parameters selected in the original paper \cite{raffel2019exploring} since we expect little margin of improvement for such a task-agnostic phase. For the fine-tuning, due to computational feasibility reasons, we did not change the model architecture (\eg number of layers), but we experiment with different learning rates schedulers. We are aware that a more extensive calibration would likely produce better results. 
\rev{Finally, we pre-trained the model by masking 15\% of tokens (\ie words in comments and code tokens in the raw and abstracted code) in the $\sim$2.7M instances from the pre-training dataset. However, we did not experiment with the model after pre-training to verify whether it actually learned the languages of interest (\ie raw source code, abstracted source code, and technical natural language). To address this limitation, we randomly selected 3k instances from the \BFmedium test set, both in their abstract and raw representation (6k in total). We also selected 3k code summaries from the \CS dataset obtaining a dataset of 9k instances, equally split across raw source code, abstracted source code, and technical natural language. Note that these are instances that have not been used to pre-train the model and, thus, are unseen for a model only subject to pre-training. We randomly masked 15\% of tokens in each of those instances, asking the pre-trained model to predict them. T5 correctly predicted 87,139 out of the 102,711 masked tokens (\ie 84.8\% accuracy). As expected, given the different complexity of the three ``languages'', T5 achieved a higher accuracy of 90.8\% when working on abstracted code, 82.7\% on raw code, and 64.6\% when guessing tokens from technical language. Overall, such results indicate that the model successfully gathered knowledge about the languages of interest during the pre-training.} 

Also the quality of the employed datasets can dramatically impact the achieved results. This is because there may be biases making the dataset not representative of the real world. To assess the quality of our datasets we conducted various analyses around sampling bias and data snooping as recommended by Watson \etal \cite{watson2020systematic}.

To this end, we conducted an exploratory data analysis (EDA), which helps answering questions related to the reliability and quality of our datasets. To accomplish this, we performed a two-fold statistical procedure: complexity size and token distributions. In the complexity size procedure, we count the number of tokens per dataset and data partition. Then, we present the relative distribution in log scale. While in the token procedure, we concentrated on counting specific tokens by popularity or special interest (\eg $if$, $assert$, or $public$). The purpose of the EDA is to monitor the size of datasets and its impact in the model performance. EDA's results can be found in our web appendix \cite{replication}. 

\textbf{Conclusion validity.} Threats to conclusion validity concern the relationship between evaluation and outcome. To this extent, we used appropriate statistical procedures, also adopting $p$-value adjustment when multiple tests were used within the same analysis.

\textbf{External validity.} Threats to external validity are related to the generalizability of our findings. Our study focused on the T5 model on four tasks using six datasets, all of which only involved Java code. While it is unclear how our model would perform if trained on other programming languages, excluding the abstraction component, the whole pipeline is language agnostic and can be easily adapted to other languages for evaluating this.

We also performed an analysis of our dataset aimed at finding out the generalizability of our models. This analysis assessed the level of data snooping among our datasets' training and test sets and how this impacts our model's results. To accomplish this we calculate the overlap between our fine-tuning datasets' training and test sets by computing the pairwise Levenshtein Distance \cite{levenshtein1966} between the two sets. With these distances calculated, we computed the correlation between the distances and the performance of our model on the different test sets.

Specifically, we selected a statistically representative sample (confidence level = $95\%$ and confidence interval = $5\%$) of each training set and calculated the pairwise Levenshtein Distance \cite{levenshtein1966} between it and the entirety of the test set for each fine-tuning dataset. Next, depending on the type of performance metric (Perfect Prediction or BLEU Score), we calculate the correlation between the minimum, median, and maximum distances of all sampled training examples to each test example and the performance of our model on the test set. For the perfect prediction, we use Point Biserial Correlation (PBC) \cite{tate1954correlation} as it allows to compare binary and continuous data. For the BLEU score, we use Pearson Correlation \cite{tate1954correlation} since both are continuous values.

\begin{table}[h]
\centering
	\caption{Correlation between training-test set similarity and test set performance.}
	\label{tbl:external-validity}
		\begin{tabular}{l r r r}
			\toprule
			Dataset  & Min & Median & Max\\ 			\midrule                        
			\BFsmall   & -0.15                     & -0.03          & 0.04 \\
			\BFmedium   & -0.05                     & -0.03          & 0.01 \\
			\MGident  & 0.21 & 0.03           & -0.23 \\
			\AGabs & -0.21                     & -0.14          & 0.29 \\
			\AGraw  & -0.21                     & -0.14          & 0.19 \\
			\CS & -0.38            & -0.17 & -0.09\\
			\bottomrule
		\end{tabular}
\end{table}

\tabref{tbl:external-validity} shows the correlation for each dataset. As shown, there exists a negative correlation between the minimum and median distances and model performance, \ie the model tends to perform worse as the distance between the training and test examples increases. For the maximum distance case, there is instead a positive correlation for perfect prediction performance, \ie the model tends to perform better the further away the maximum training examples are from the test examples. Such a result may be simply due to specific outliers present in the test set (\ie an instances being very far from the ones in the training set). However, all the correlations we observed are quite low, supporting the generalizability of our models. 

\section{Conclusion} \label{sec:conclusion}
We presented an empirical study aimed at investigating the usage of transfer learning for code-related tasks. In particular, we pre-trained and fine-tuned several variants of the Text-To-Text Transfer Transformer (T5) model with the goal of supporting four code-related tasks, namely \emph{automatic bug-fixing}, \emph{injection of code mutants}, \emph{generation of assert statements in test methods}, and \emph{code summarization}. We compared the performance achieved by the T5 against state-of-the-art baselines that proposed DL-based solutions to these four tasks.

The achieved results showed that: (i) the pre-training process of the T5, as expected, boosts its performance across all tasks; (ii) the multi-task fine-tuning (\ie a single model trained for different tasks) instead, does not consistently help in improving performance, possibly due to the different types of ``data'' manipulated in the four tasks (\ie raw code, abstracted code, natural language); (iii) in its best configuration, the T5 performs better that the baselines across all four tasks.
\rev{When looking at the latter finding it is important to remember that the baselines used for comparison are not pre-trained and, thus, they (i) exploited less training data, and (ii) did not need the additional $\sim$22 hours of computation required by the pre-training. 
}

Future work will aim at further advancing performance by employing larger versions of the T5. Also, while our results do not support the usage of multi-task learning in code-related tasks, we believe additional investigations are needed on this side. For example, by only considering a set of tasks all manipulating the same type of data (\eg all working on raw code), it is possible that the benefits of multi-task learning would emerge.

\section*{Acknowledgment}
This project has received funding from the European Research Council (ERC) under the European Union's Horizon 2020 research and innovation programme (grant agreement No. 851720). W\&M team has been supported in part by the NSF CCF-1955853, CCF-1815186 and CCF-2007246 grants. Any opinions, findings, and conclusions expressed herein are the authors' and do not necessarily reflect those of the sponsors. 

\balance

\bibliographystyle{IEEEtranS}
\bibliography{main}
\newpage

\begin{IEEEbiography}[{\includegraphics[width=1in,height=1.25in,clip,keepaspectratio]{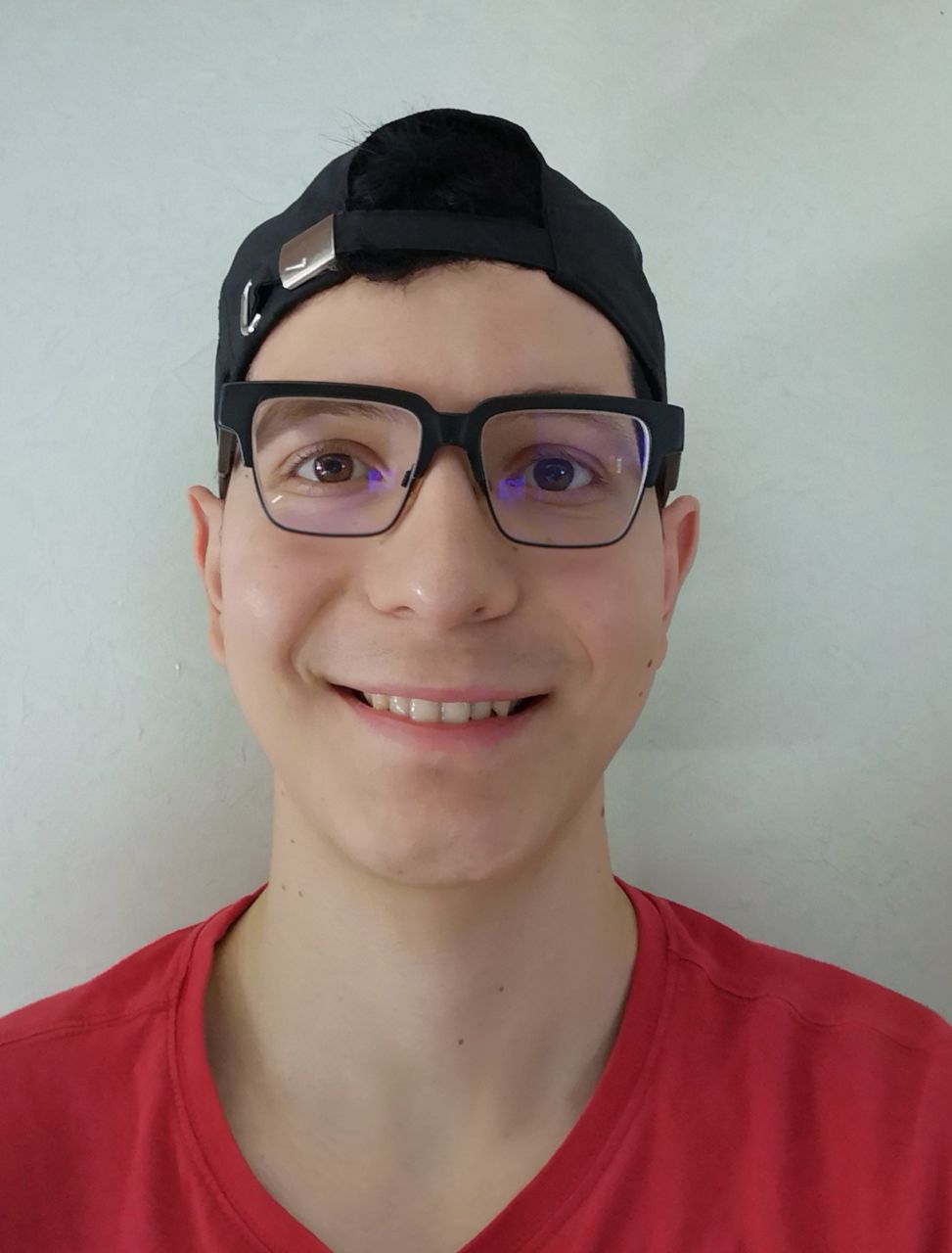}}]{Antonio Mastropaolo} is a Ph.D. student in the Faculty of Informatics at the Universit\`a della Svizzera italiana (USI), Switzerland, where he is part of the Software Institute. He received his MSc. in Software System Security from Universit\`a degli studi del Molise, Italy, in July 2020. His research interests include the study and the application of deep-learning techniques to foster code-related tasks. More information available at: \url{https://antoniomastropaolo.com}.
\end{IEEEbiography}

\begin{IEEEbiography}[{\includegraphics[width=1in,height=1.25in,clip,keepaspectratio]{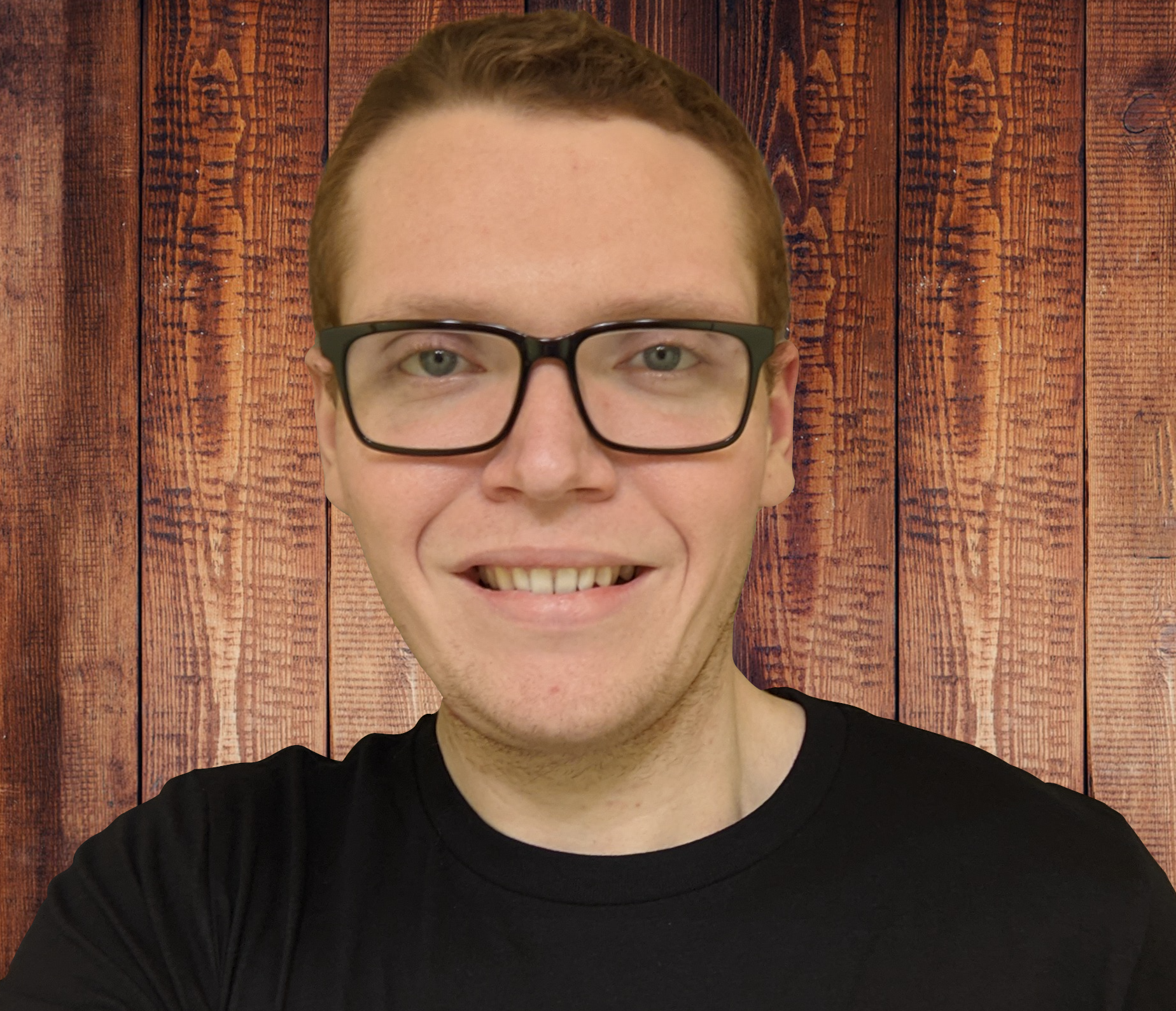}}]{Nathan Cooper} received a B.S. degree in Software Engineering from the University of West Florida in 2018. He is currently a Ph.D. candidate in Computer Science at William \& Mary under the advisement of Dr. Denys Poshyvanyk and is a member of the Semeru Research group. He has research interests in Software Engineering, Machine / Deep Learning applications for Software Engineering, information retrieval, and question \& answering applications for Software Engineering. He has published in the top peer-reviewed Software Engineering venues ICSE and MSR. He has also received the ACM SIGSOFT Distinguished paper award at ICSE'20. More information is available at \url{https://nathancooper.io/#/}.
\end{IEEEbiography}

\begin{IEEEbiography}[{\includegraphics[width=1in,height=1.25in,clip,keepaspectratio]{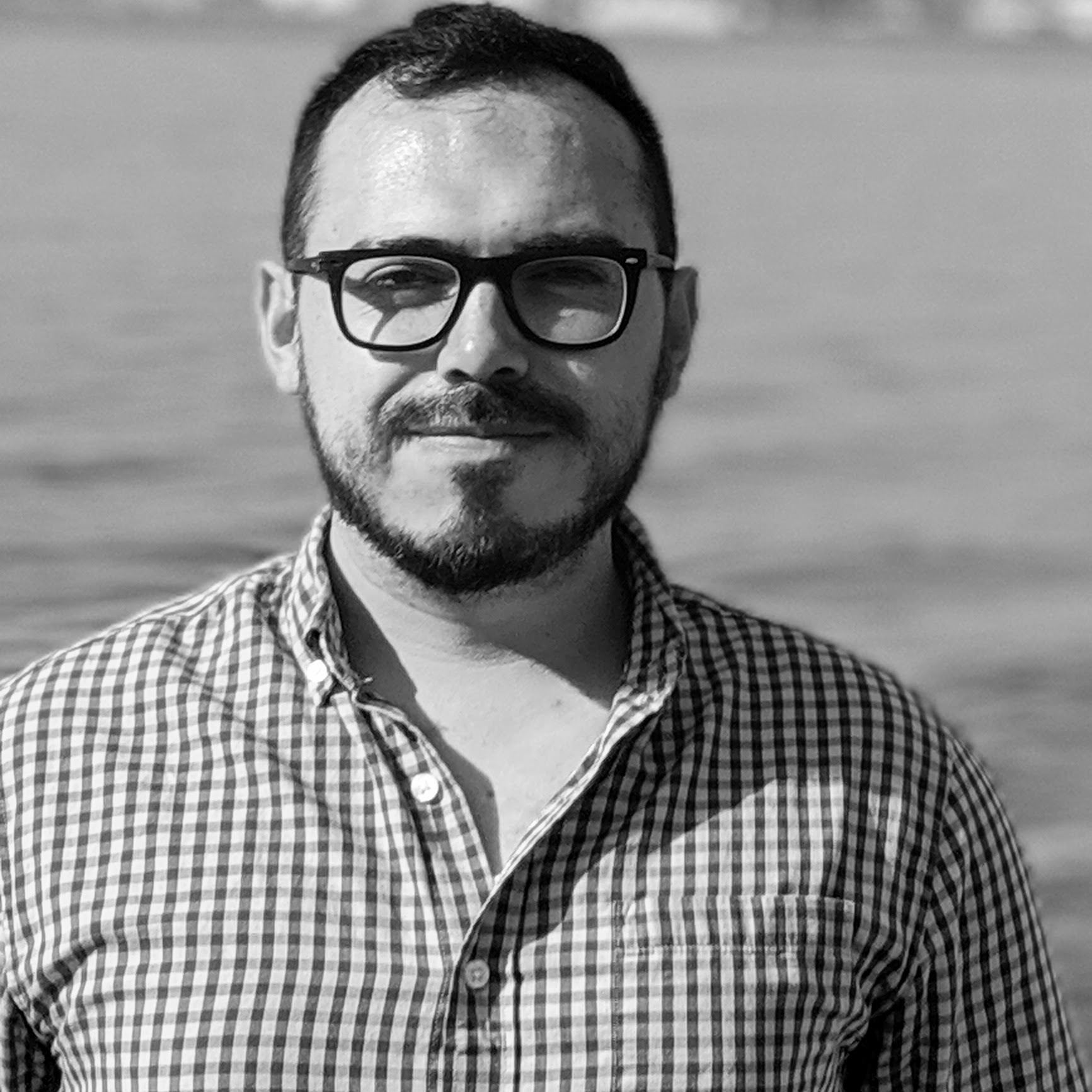}}]{David N. Palacio} is a Ph.D. Candidate in Computer Science at The College of William \& Mary, where he is a member of the SEMERU Research Group supervised by Dr. Denys Poshyvanyk. He received his MSc. in Computer Engineering at Universidad Nacional de Colombia (UNAL), Colombia, 2017. His research is concentrated on interpretable methods for deep learning code generators, specifically, towards using causal inference to explain deep software models. His fields of interest lie in complexity science, neuroevolution, causal inference, and interpretable machine learning for the study and automation of software engineer processes. More information available at \url{https://danaderp.github.io/danaderp/}.
\end{IEEEbiography}

\begin{IEEEbiography}[{\includegraphics[width=1in,height=1.25in,clip,keepaspectratio]{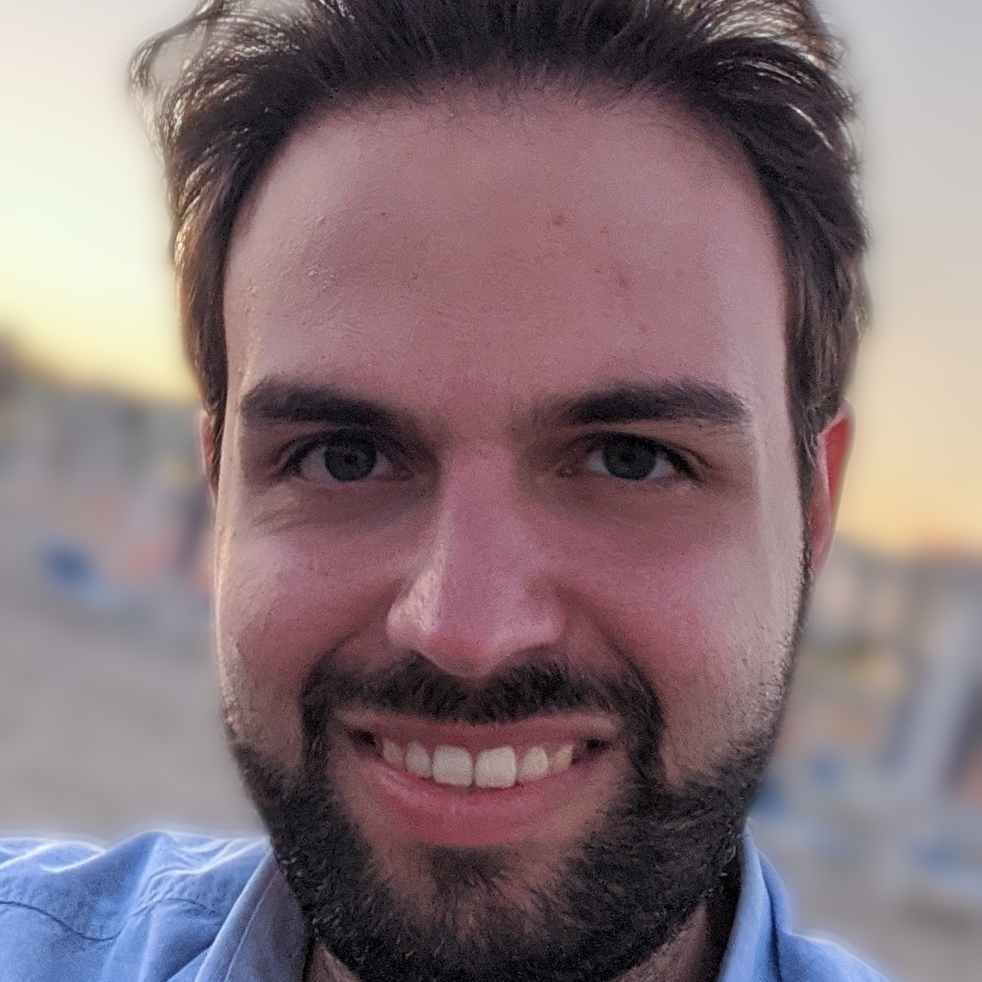}}]{Simone Scalabrino} is a Research Fellow at the University of Molise, Italy. He has received his MS degree from the University of Salerno, and his PhD degree from the University of Molise, defending a thesis on automatically assessing and improving source code readability and understandability. His main research interests include code quality, software testing, and empirical software engineering. He has received three ACM SIGSOFT Distinguished Paper Awards at ICPC 2016, ASE 2017, and MSR 2019. He is co-founder and CSO of datasound, a spin-off of the University of Molise.
More information available at: \url{https://dibt.unimol.it/sscalabrino/}.
\end{IEEEbiography}

\begin{IEEEbiography}[{\includegraphics[width=1in,height=1.25in,clip,keepaspectratio]{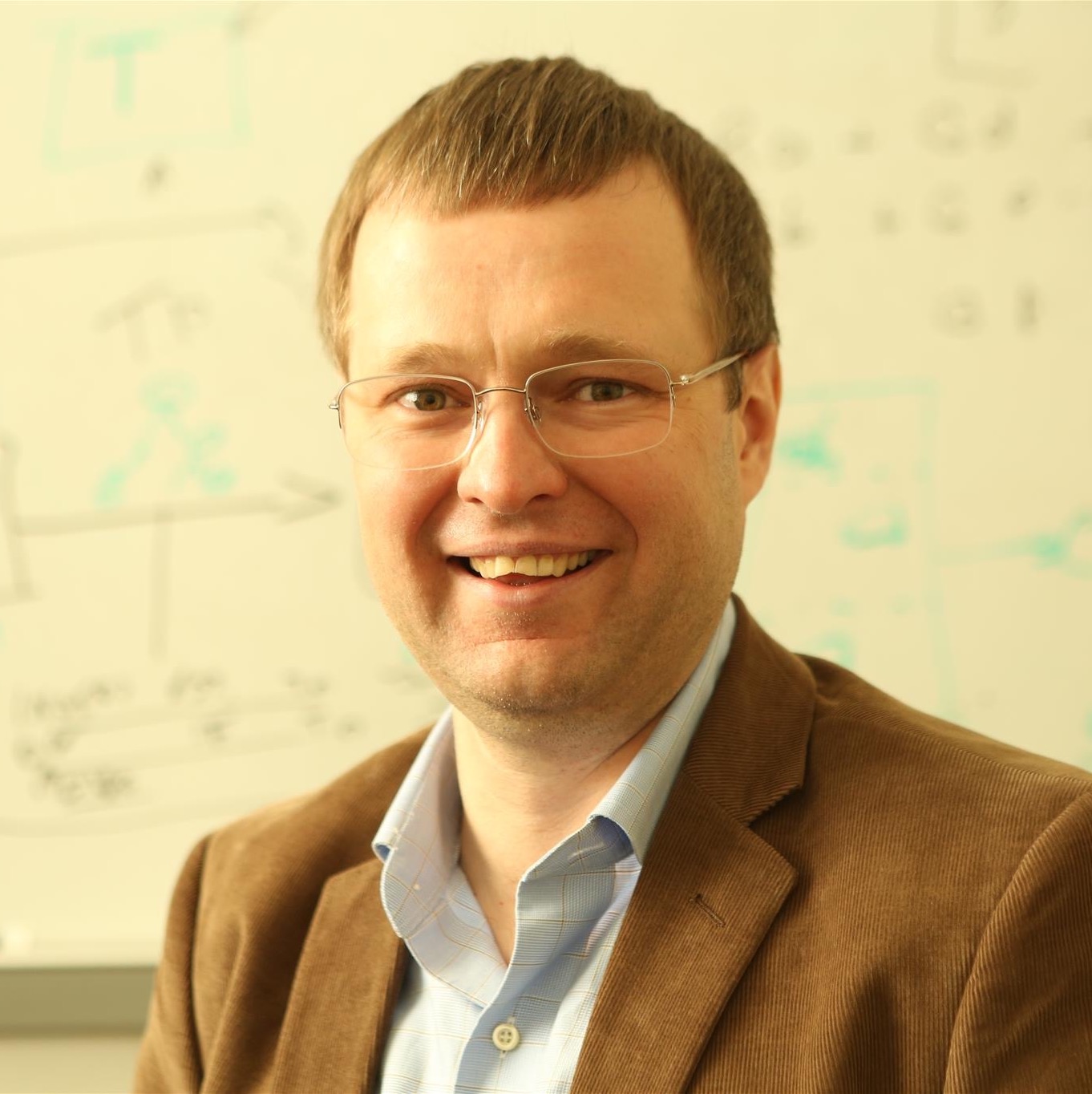}}]{Denys Poshyvanyk} is a Professor of Computer Science at William and Mary. He received the MS and MA degrees in Computer Science from the National University of Kyiv-Mohyla Academy, Ukraine, and Wayne State University in 2003 and 2006, respectively. He received the PhD degree in Computer Science from Wayne State University in 2008. He served as a program co-chair for ASE'21, MobileSoft'19, ICSME'16, ICPC'13, WCRE'12 and WCRE'11. He currently serves on the editorial board of IEEE Transactions on Software Engineering (TSE), ACM Transactions on Software Engineering and Methodology (TOSEM), Empirical Software Engineering Journal (EMSE, Springer), Journal of Software: Evolution and Process (JSEP, Wiley) and Science of Computer Programming. His research interests include software engineering, software maintenance and evolution, program comprehension, reverse engineering and software repository mining. His research papers received several Best Paper Awards at ICPC'06, ICPC'07, ICSM'10, SCAM'10, ICSM'13, CODAPSY'19 and ACM SIGSOFT Distinguished Paper Awards at ASE'13, ICSE'15, ESEC/FSE'15, ICPC'16, ASE'17, ESEC/FSE'19 and ICSE'20. He also received the Most Influential Paper Awards at ICSME'16, ICPC'17, ICPC'20 and ICSME'21. He is a recipient of the NSF CAREER award (2013).  He is a member of the IEEE and ACM. More information is available at: \url{http://www.cs.wm.edu/~denys/}.
\end{IEEEbiography}

\begin{IEEEbiography}[{\includegraphics[width=1in,height=1.25in,clip,keepaspectratio]{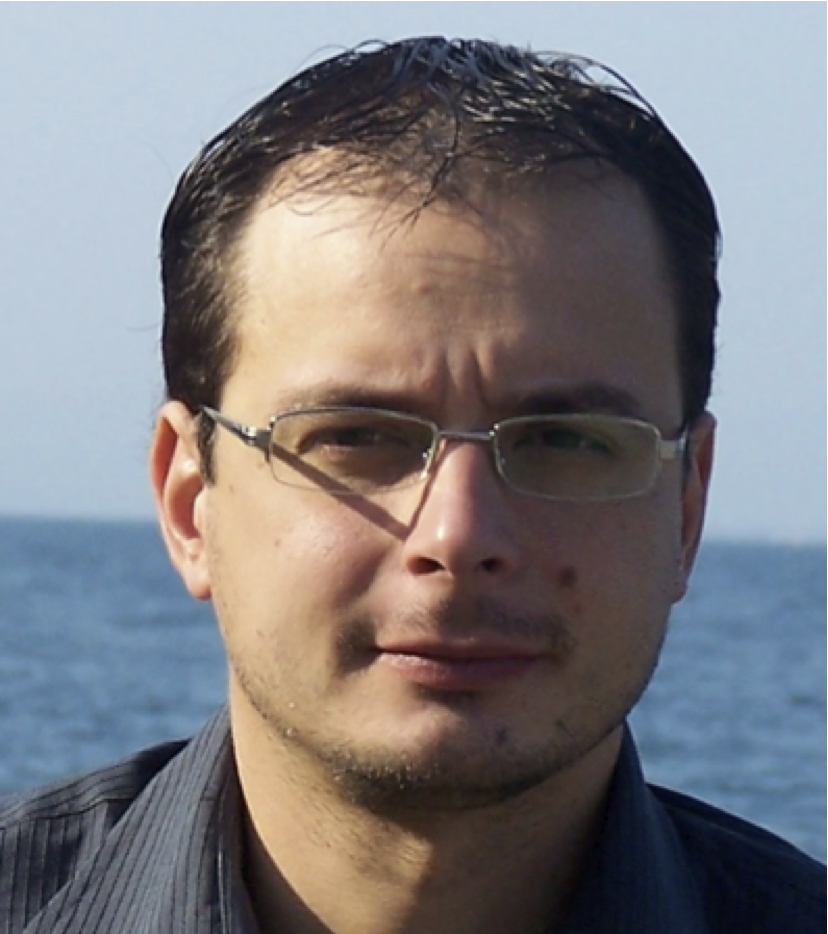}}]{Rocco Oliveto} is a Professor in the Department of Bioscience and Territory at University of Molise (Italy). He is the Chair of the Computer Science program and the Director of the Laboratory of Computer Science and Scientific Computation of the University of Molise. He received the PhD in Computer Science from University of Salerno (Italy) in 2008. His research interests include traceability management, information retrieval, software maintenance and evolution, search-based software engineering, and empirical software engineering. He is author of about 150 papers appeared in international journals, conferences and workshops. He serves and has served as organizing and program committee member of international conferences in the field of software engineering. He is a member of IEEE Computer Society and ACM.
\end{IEEEbiography}

\begin{IEEEbiography}[{\includegraphics[width=1in,height=1.25in,clip,keepaspectratio]{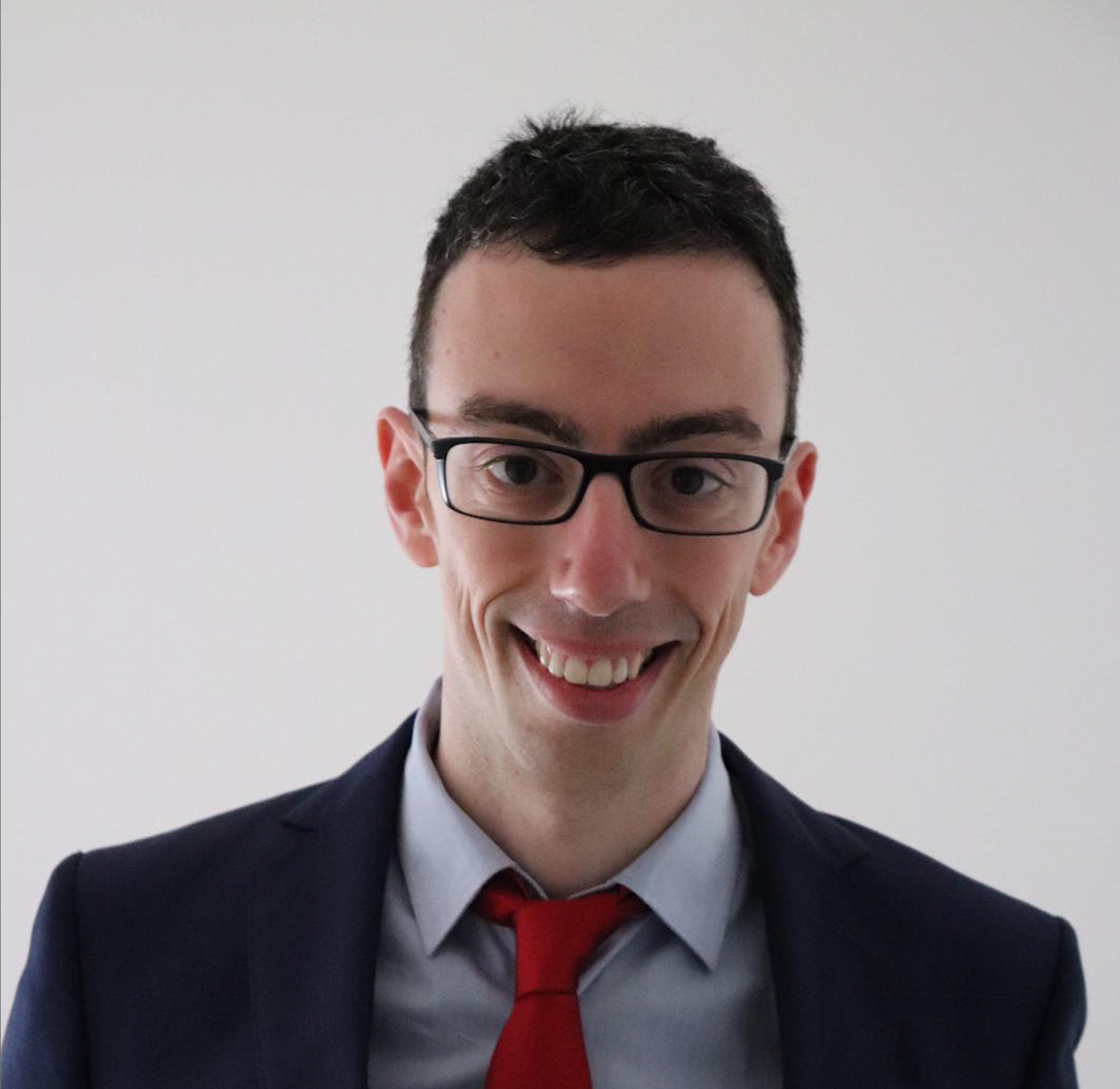}}]{Gabriele Bavota} is an associate professor at the Faculty of Informatics of the Universit\`a della Svizzera italiana (USI), Switzerland, where he is part of the Software Institute and he leads the SEART research group. He received the PhD in Computer Science from the University of Salerno, Italy, in 2013. His research interests include software maintenance and evolution, code quality, mining software repositories, and empirical software engineering. On these topics, he authored over 140 papers appeared in international journals and conferences and has received four ACM Sigsoft Distinguished Paper awards at the three top software engineering conferences: ASE 2013 and 2017, ESEC-FSE 2015, and ICSE 2015. He also received the best/distinguished paper award at SCAM 2012, ICSME 2018, MSR 2019, and ICPC 2020.
He is the recipient of the 2018 ACM Sigsoft Early Career Researcher Award for outstanding contributions in the area of software engineering as an early career investigator and the principal investigator of the DEVINTA ERC project. More information is available at: \url{https://www.inf.usi.ch/faculty/bavota/}.
\end{IEEEbiography}

\end{document}